\documentclass[aip,jmp]{revtex4-1}
\usepackage{slashed}
\usepackage{amsmath}
\usepackage{amssymb}

\begin{document}

\title{Breaking Generalized Covariance, Classical Renormalization and Boundary Conditions from Superpotentials}

\author{Gideon I.~Livshits}
\email[]{livshits.gideon@mail.huji.ac.il}

\affiliation{Hebrew University of Jerusalem, Institute of Chemistry, Edmond J. Safra Campus, Jerusalem, Israel 91904}

\begin{abstract}
Superpotentials offer a direct means of calculating conserved charges associated with the asymptotic symmetries of space-time. Yet superpotentials have been plagued with inconsistencies, resulting in nonphysical or incongruent values for the mass, angular momentum and energy loss due to radiation. The approach of Regge and Teitelboim, aimed at a clear Hamiltonian formulation with a boundary, and its extension to the Lagrangian formulation by Julia and Silva have resolved these issues, and have resulted in a consistent, well-defined and unique variational equation for the superpotential, thereby placing it on a firm footing. A hallmark solution of this equation is the KBL superpotential obtained from the first-order Lovelock Lagrangian. Nevertheless, here we show that these formulations are still insufficient for Lovelock Lagrangians of higher orders. We present a paradox, whereby the choice of fields affects the superpotential for equivalent on-shell dynamics. We offer two solutions to this paradox: either the original Lagrangian must be effectively renormalized, or that boundary conditions must be imposed, so that space-time be asymptotically maximally symmetric. Non-metricity is central to this paradox, and we show how quadratic non-metricity in the bulk of space-time contributes to the conserved charges on the boundary, where it vanishes identically. This is a realization of the gravitational Higgs mechanism, proposed by Percacci, where the non-metricity is the analogue of the Goldstone boson.
\end{abstract}


\maketitle 

\section{\label{sec:level1}Introduction}
A space-time with the most general distribution of matter and energy is a space-time for which it is, \textit{a priori}, impossible to define globally conserved charges \cite{Ge,Wa}. However, when the sources of gravity are isolated and localized to a specific region in space-time, one may, if sufficiently far away from the sources, define unique integration at infinity to obtain values for the globally conserved mass or energy, angular momentum or energy flux (in the case of isolated sources of gravitational radiation). This is the working premise behind the use of superpotentials to calculate conserved charges \cite{Pa,Katz3}. They are directly related to the differential conservation laws associated with the asymptotic symmetries of space-time. Integrated on the relevant slice at the appropriate infinity, superpotentials can provide the mass of an isolated source of gravity, its angular momentum or, if it is radiating, its energy loss.

For decades superpotentials were plagued with inconsistencies and were a key source of problems in gravitational theory, the most striking of which were their non-covariance and their non-uniqueness \cite{Katz3}. Non-covariance implies that charges depend on a specific choice of coordinate frames, in clear violation of the principle of covariance, which is a corner stone of general relativity (GR). Non-uniqueness implies two deficiencies: the first, that the same action could produce different superpotentials; and the second, that different superpotentials had to be constructed for different space-time geometries. While a specific superpotential would yield the expected result for one type of charge for a given geometry, it would give the wrong result when evaluated for a different charge. The pathologies \cite{Misner1963} and the famous \textquotedblleft factor $2$\textquotedblright\ problem \cite{K1985} of the Komar superpotential \cite{Ko} are just one such example. The literature on this subject is quite vast, and no single work encompasses all the different currents and superpotentials which have been devised to solve specific problems. A recent review by Szabados \cite{Sz} contains many examples, but is not exhaustive. See also extensive bibliography in a recent series of publications by Lompay and Petrov \cite{LP1,LP2}.

These inconsistencies with the general principles of GR implied there was something inherently wrong with the way these conserved charges were defined and calculated, and with superpotentials in particular. It was unclear whether a single superpotential could be formulated which would work for all geometries of isolated sources of gravity, until in $1985$ Katz proposed what would become known as the KBL superpotential \cite{K1985,KBL}.

This superpotential was shown to solve these two problems by introducing a background space-time to solve the problem of non-covariance and by implementing a well-defined variational principle to solve the problem of non-uniqueness \cite{RT}.

The background in this approach is fixed and used only on the boundary of space-time to eliminate the degeneracy in the choice of a coordinate system, which produces non-physical infinities. The removal of the background quantity in the integration has a similar effect to that of renormalization in quantum field theory (QFT).

The problem of non-uniqueness is partially solved simply by the fact that for any known geometry of isolated sources of gravity, whether at spatial or null infinity, whether in $N=4$ or more dimensions, the KBL superpotential provides the correct mass, angular momentum or radiative energy loss. In the original paper \cite{K1985}, it was shown to give the correct Bondi mass loss \cite{Bo} in four dimensions, and more recently, in five dimensions as well \cite{KL2}. It is the only superpotential which possesses all the following properties. It is generally covariant and can be computed in any coordinate system. In Cartesian coordinates of an asymptotically flat space-time it gives the ADM mass formula \cite{ADM} and in asymptotically anti-de Sitter space-times it gives the AD mass \cite{AD}. It gives the mass and angular momentum as well as the Brown and Henneaux conformal charges \cite{BH} with the right normalization in any dimension, $N\geq 3$. It can also be used for any Killing vector field of the background. It reproduces the Penrose mass \cite{Pen1}, the Penrose linear momentum \cite{Pen1}, the Sachs linear momentum flux \cite{Sachs} and the Penrose \cite{Pen2} and Dray and Streubel \cite{DS} angular momentum at null infinity. In addition to these, it gives the mass of a Kerr black hole in an anti-de Sitter background in $N$ dimensions \cite{DK1}. In summary, it requires no amendment nor any additional artificial terms for specific geometries.

That is only one part of the uniqueness question. The other part is fulfilled by the implementation of the Regge-Teitelboim approach, originally developed for the well-posedness of the gravitational Hamiltonian in the presence of boundaries \cite{RT}. Julia and Silva were the first to implement this approach in the Lagrangian formulation by providing a uniqueness criterion \cite{JS3,JS2,S1,JS1}. This criterion amounts to a variational differential equation that determines the variation of the superpotential via variational derivatives, and which stems from the physical requirement that the divergence terms, present in the variation of the conserved current, vanish on the boundary. Coupled with appropriate boundary conditions at spatial infinity, it reproduced the KBL superpotential for the first-order Lovelock action \cite{JS1}. The arbitrariness of divergence terms that could be added to the action is removed, since variational derivatives are immune to such additions.

In that paper and throughout their work, Julia and Silva worked within the metric-affine or general linear, $GL\left(N,\mathbf{R}\right)$, formulation of relativity \cite{HEHL1}. This is the most general first-order formulation of gravity, in which the variations of the fields and the field equations contain at most first-order derivatives of the fields / parameters. This feature of the $GL\left(N,\mathbf{R}\right)$ formulation makes it so effectively simple to handle. We extended their criterion to the Palatini formulation \cite{KL1}, and provided an extension of their equation for the case where field equations were still first-order, but the variations contained second-order derivatives of the symmetry parameter, which was now reduced only to diffeomorphism invariance.

In our previous studies \cite{KL1,KL2} we solved the equations in both formulations for the first and second-order Lovelock Lagrangians. In the former case, both solutions reproduced the KBL superpotential exactly. However, starting with the second-order Lovelock Lagrangian, the solutions diverge. In this paper, we compare the results of both the $GL\left(N,\mathbf{R}\right)$ and the Palatini formulations and their respective equations, and explore this profound difference. We find that even though they stem from the same uniqueness criterion, and are derived from equivalent currents and subjected to equivalent boundary conditions (save for gauge fixing and symmetry breaking) with \textbf{the same on-shell dynamics}, the superpotentials depend on the representation of the theory. That is, the representation - the choice of dynamic fields and their associated symmetries - has a direct measurable effect on the physical charge that is calculated at the boundary. In this sense we say that \emph{generalized covariance} is broken. Further still, if we choose to maintain generalized covariance, then this imposes new conditions: either we must renormalize the Palatini Lagrangian by adding to it a new Lagrangian density, which artificially removes the difference, or we must impose constraints on the asymptotic structure of space-time.

We show that the non-metricity tensor plays a key role in this discrepancy, even though it and its higher derivative terms vanish on the boundary. We construct different actions, both parity-preserving and parity-violating, which contain explicit quadratic non-metricity. The vanishing of non-metricity on the boundary and its dynamical effect on curvature is reminiscent of the gravitational Higgs mechanism \cite{P}, proposed by Percacci in the mid-1980's. We examine the contribution of non-metricity to the conserved charges on the boundary, showing that non-metricity can be utilized to test the possible breaking of parity, by its global and measurable effects at the boundary. We tie together the effects of quadratic non-metricity with the surface term, proposed by Obukhov \cite{Obukhov} for manifolds with a boundary, by showing that the renormalized first-order Lovelock Lagrangian fixes the variation of the boundary term.

In Sec. II we briefly outline the elements of the metric-affine formulation, the Palatini formulation, the work of Julia and Silva and the extension of their equations to second-order derivatives of the symmetry parameter. In Sec. III we solve both sets of equations for each formulation of GR, and show how the superpotentials differ for higher-order curvature terms. In Sec. IV we show how to renormalize Lovelock Lagrangians, and provide numerous examples of Lagrangian densities that couple non-metricity to the curvature which can generate superpotentials. As an example application, a parity-violating Lagrangian density is shown to induce a superpotential in a maximally symmetric background. The conserved charges are calculated for an asymptotically Kerr-AdS metric in four dimensions \cite{HT} and a gravo-magnetic analogue based on the Taub-NUT solution in four dimensions \cite{LBNZ}. Finally, section V discusses and summarizes the source of the difference.

\section{\label{sec:level2}Elements}

Superpotentials constructed by the standard Noether method (such as the Komar superpotential \cite{Katz3,Ko,DK2}) lack uniqueness. Any divergence that may be added to the Lagrangian density can affect the final outcome, while the dynamics are unaltered. In fact, it was just such a divergence that was added in the original derivation of the KBL superpotential \cite{K1985,KBL}. In a series of papers, Julia and Silva \cite{JS1,JS2,S1,JS3} studied covariant phase space methods, and realized that the superpotential could be constructed out of variational derivatives, making it invariant to these divergences. In this way the physical charge associated with symmetries of the action could be unambiguously defined on the boundary.

As they argue themselves, their method is nothing more than a generalization to the Lagrangian formulation of the work of Regge and Teitelboim \cite{RT}, which was originally proposed to make the Hamiltonian formulation well-defined when boundaries were present. Intuitively \cite{KL2}, the variation of the Hamiltonian should have no boundary terms by analogy with classical mechanics. For a more complete treatment of this matter the reader is advised to revisit \cite{KL2} and references therein.

Their original prescription \cite{JS1} applies to an entirely first-order theory, in the sense that:
\begin{itemize}
\item variational derivatives depend at most on first-order derivatives of the field components;

\item variations of these field components with respect to the symmetry parameters (so-called \emph{variational symmetries}) must also contain at most first-order derivatives of the fields
and the symmetry parameters.
\end{itemize}

These two conditions are met in the most general first-order formulation of Einstein's theory, the so-called metric-affine gravity, or $GL\left( N,\mathbf{R}\right)$ formulation. In fact, both conditions are satisfied by all Lovelock Lagrangians within this formulation. This very
fact provides the motivation to use this formulation over other formulations, such as the Palatini formulation, in which the second
condition is not met, or the original metric formulation in which the first condition is not satisfied.

The basics are presented below in the language of components, which facilitates the comparison of the results of the $GL\left(N,\mathbf{R}\right)$ and the Palatini formulations.

\subsection{Elements of the $GL\left( N,\mathbf{R}\right)$ formulation}

In the $GL\left( N,\mathbf{R}\right) $ formulation the dynamic fields are the fibre metric $\gamma^{ab}$, the non-coordinate base (or soldering form) $\theta_{~\mu }^{a}$, and the spin connection, $\omega _{\mu ~b}^{~a}$. They are taken to be independent of each other. The $\theta_{~\mu }^{a}$'s are assumed invertible, obeying
\begin{equation}
\theta _{~\mu }^{a}\left( x\right) \theta _{b}^{~\mu }\left(x\right)=\delta_{b}^{a}\text{ \ and\ \ }\theta _{a}^{~\mu }\left( x\right)
\theta_{~\nu}^{a}\left( x\right) =\delta _{\nu }^{\mu }.
\label{1}
\end{equation}%
The fibre metric is related to the space-time metric via
\begin{equation}
\gamma^{ab}=\theta_{~\alpha }^{a}\theta_{~\beta }^{b}g^{\alpha \beta }.
\label{2}
\end{equation}%
The first and second-order actions are given by%
\begin{equation}
I_{n}\left[ \gamma^{ab},\theta _{~\mu }^{a},\omega _{\mu ~b}^{~a}\right]=\alpha _{n}\int \limits_\mathcal{M} \! \hat{\mathcal{L}}_{n} \, \mathrm{d}V,~n=1,2  \label{3}
\end{equation}%
where $\mathcal{M}$ is the $N$-dimensional physical space-time with boundary $\partial\mathcal{M}$, a caret denotes a density of weight $+1$, and $\hat{\mathcal{L}}_{1}$, $\hat{\mathcal{L}}_{2}$ are the first and second-order Lovelock Lagrangian densities:%
\begin{equation}
\hat{\mathcal{L}}_{1} = \frac{1}{2}\sqrt{\left\vert \gamma \right\vert }\gamma^{cb}\theta_{ac}^{\mu \nu }B_{~b\mu \nu }^{a}; ~\hat{\mathcal{L}}_{2} = \frac{1}{4}\sqrt{\left\vert \gamma \right\vert }\gamma^{mb}\gamma^{nd}\theta_{amcn}^{\mu \nu \rho \sigma }B_{~b\mu\nu}^{a}B_{~d\rho \sigma }^{c},
\label{4}
\end{equation}%
with%
\begin{equation}
\theta _{ab}^{\mu \nu }=-\frac{\left( -1\right) ^{N}}{\left( N-2\right) !}
\epsilon_{aba_{1}\ldots a_{N-2}}\epsilon ^{\mu \nu \alpha _{1}\ldots \alpha
_{N-2}} \theta _{~~\alpha _{1}}^{a_{1}}\theta
_{~~\alpha_{2}}^{a_{2}}\cdots \theta _{~~~~~~\alpha _{N-2}}^{a_{N-2}};
\label{6}
\end{equation}%
\begin{equation}
\theta _{abcd}^{\kappa \lambda \mu \nu }=-\frac{\left( -1\right) ^{N}}{
\left( N-4\right) !}\epsilon _{abcda_{1}\ldots a_{N-4}}\epsilon ^{\kappa
\lambda \mu \nu \alpha _{1}\ldots \alpha _{N-4}} \theta _{~~\alpha _{1}}^{a_{1}}\theta
_{~~\alpha_{2}}^{a_{2}}\cdots \theta _{~~~~~~\alpha _{N-4}}^{a_{N-4}}.
\label{7}
\end{equation}%
The physical units are preserved in the coefficients $\alpha_{n}$. Lower
case Latin indices, $a$, $b$,..., denote group indices, while Greek indices,
$\mu$, $\nu$,..., denote space-time indices. $\epsilon _{a_{1}a_{2}\ldots
a_{N}}$ is the totally anti-symmetric tensor with $\epsilon_{12\cdots N}=+1$%
. The Riemann curvature tensor is a function of the spin connection and its
first-order derivatives:
\begin{equation}
B_{~b\mu \nu }^{a}\left[ \mbox{\boldmath$\omega$}\right] =2\partial _{\left[
\mu\right.}\omega _{\left. \nu \right] ~b}^{~~a}+2\omega _{\left[
\mu\right\vert ~c}^{~~a}\omega_{\left\vert \nu \right] ~b}^{~~c}.  \label{8}
\end{equation}%
Square brackets denote antisymmetrization, as $X_{\left[\mu \nu \right]}
\equiv \frac{1}{2} (X_{\mu \nu} - X_{\nu \mu})$. We assume at the outset the
vanishing of torsion, which is given by
\begin{equation}
\nabla _{\left[ \kappa \right. }\theta _{~\left. \lambda \right] }^{a}=0%
\text{ or }\Gamma _{~\left[ \mu \nu \right] }^{\lambda }=0:\text{%
torsion-free condition,}  \label{9}
\end{equation}%
where, in going from the first to the second equation in (\ref{9}) we have
implicitly assumed the vielbein postulate. This postulate relates the spin
connection to the affine connection, $\Gamma_{~\mu \nu}^{\lambda}$,  via
\begin{equation}
\omega _{\mu ~b}^{~a}=\theta _{b}^{~\lambda }\left( \theta_{~\nu
}^{a}\Gamma_{~\mu \lambda }^{\nu}-\partial _{\mu }\theta
_{~\lambda}^{a}\right) =\theta_{~\lambda }^{a}D_{\mu }\theta _{b}^{~\lambda
},  \label{10}
\end{equation}%
where $D_{\mu}$ is the covariant derivative with respect to $\Gamma_{~\mu \nu}^{\lambda}$ acting on space-time indices, e.g.
\begin{equation}
D_{\mu }V_{b}^{~\lambda }=\partial _{\mu
}V_{b}^{~\lambda}+\Gamma_{~\mu\nu}^{\lambda }V_{b}^{~\nu
},~D_{\mu}V_{~\lambda }^{b}=\partial _{\mu}V_{~\lambda }^{b}-\Gamma _{~\mu
\lambda }^{\nu }V_{~\nu }^{b}\text{, etc.}
\label{11}
\end{equation}%
We must stress here that we shall use (\ref{10}) \emph{explicitly} only on the boundary when breaking symmetry by going from $GL\left(N,\mathbf{R}\right)$ to Palatini (see more in section II.D). From eqs.~(\ref{8}) and (\ref{10}) it further follows that
\begin{equation}
B_{~b\mu \nu }^{a}\left[ \mbox{\boldmath$\omega$} \right] =\theta_{~\rho}^{a}\theta_{b}^{~\sigma }B_{~\sigma \mu \nu }^{\rho }\left[\mathbf{\Gamma}\right]
\label{12}
\end{equation}
with
\begin{equation}
B_{~\sigma \mu \nu }^{\rho }\left[ \mathbf{\Gamma} \right] =2\partial _{\left[%
\mu\right. }\Gamma _{~\left. \nu \right] \sigma }^{\rho }+2\Gamma _{~\left[
\mu\right\vert\lambda }^{\rho }\Gamma _{~\left\vert \nu \right]%
\sigma}^{\lambda }.
\label{13}
\end{equation}%
$\nabla_{\kappa}$ is the covariant derivative with respect to spin
connection acting on the group indices, e.g.%
\begin{equation}
\nabla _{\kappa }\theta _{~\lambda }^{a}=\partial _{\kappa }\theta_{~\lambda
}^{a}+\omega_{\kappa ~b}^{~a}\theta _{~\lambda }^{b}\text{, }\nabla _{\kappa
}\theta_{a}^{~\lambda}=\partial _{\kappa }\theta_{a}^{~\lambda }-\omega
_{\kappa~a}^{~b}\theta_{b}^{~\lambda }\text{, etc.}
\label{14}
\end{equation}%
The variational derivatives of $\hat{\mathcal{L}}_{1}$ and $\hat{\mathcal{L}}%
_{2}$ are given by%
\begin{eqnarray}
\frac{1}{\sqrt{\left\vert \gamma \right\vert }}\frac{\delta \hat{\mathcal{L}}%
_{1}}{\delta \theta _{~\eta }^{e}} & = & \frac{1}{2}\gamma ^{cb}\theta
_{ace}^{\mu \nu \eta }B_{~b\mu \nu }^{a}=\theta _{e}^{~\eta }\mathcal{L}%
_{1}+\theta_{ab}^{\eta \mu }\theta _{e}^{~\nu }B_{~~\mu \nu }^{ab};
\label{15} \\
\frac{1}{\sqrt{\left\vert \gamma \right\vert }}\frac{\delta \hat{\mathcal{L}}%
_{1}}{\delta \gamma ^{pq}} & = & -\frac{1}{2}\gamma _{pq}\mathcal{L}_{1}+\frac{1%
}{2}\theta _{ac}^{\mu \nu }\delta _{\left( p\right.
}^{c}\delta_{\left.q\right) }^{b}B_{~b\mu \nu }^{a};  \label{16} \\
\frac{\delta \hat{\mathcal{L}}_{1}}{\delta \omega _{\lambda ~b}^{~a}} &
= & -\theta _{ac}^{\kappa \lambda }\nabla _{\kappa }\left( \sqrt{%
\left\vert\gamma \right\vert }\gamma^{bc}\right) ,  \label{17}
\end{eqnarray}
and
\begin{equation}
\frac{1}{\sqrt{\left\vert \gamma \right\vert }}\frac{\delta \hat{\mathcal{L}}_{2}}{\delta \theta _{~\eta }^{e}} \equiv \mathcal{L}_{e}^{~\eta } =\frac{1}{4} \theta_{a b c d e}^{\kappa \lambda \mu \nu \eta} B_{~~\kappa\lambda}^{ab}B_{~~\mu\nu }^{cd}  =\theta _{e}^{~\eta} \mathcal{L}_{2}-\theta_{e}^{~\kappa} \theta_{abcd}^{\eta \lambda \mu \nu}B_{~~\kappa \lambda }^{ab}B_{~~\mu \nu}^{cd};
\label{18}
\end{equation}
\begin{eqnarray}
\frac{1}{\sqrt{\left\vert \gamma \right\vert }}\frac{\delta \hat{\mathcal{L}}%
_{2}}{\delta \gamma^{pq}}& \equiv \mathcal{L}_{pq} =-\frac{1}{2}\gamma _{pq}%
\mathcal{L}_{2}+\frac{1}{2}\theta _{abcd}^{\kappa \lambda
\mu\nu}\delta_{\left( p\right.}^{b}\delta _{\left. q\right) }^{m}B_{~m\kappa
\lambda }^{a}B_{~~\mu \nu }^{cd};  \label{19} \\
\frac{\delta \hat{\mathcal{L}}_{2}}{\delta \omega _{\lambda ~b}^{~a}}& \equiv \hat{
\mathcal{L}}_{~a}^{\lambda~b} = -\nabla _{\kappa }\left( \sqrt{\left\vert
\gamma \right\vert }\gamma ^{bm}\gamma^{dn}\right)\theta_{amcn}^{\kappa\lambda \mu \nu }B_{~d\mu \nu }^{c},
\label{20}
\end{eqnarray}
respectively. The torsion-free condition, eq.~(\ref{9}), was used to eliminate torsion terms in the right hand-side of eqs.~(\ref{17}) and (\ref{20}). These variational derivatives obey the generalized Bianchi identity (also known as the $2^{\text{nd}}$ Noether identity \cite{HEHL1}):
\begin{equation}
\nabla _{\lambda }\left( \frac{\delta \hat{\mathcal{L}}}{\delta
\omega_{\lambda ~b}^{~a}}\right) +\theta _{~\eta }^{b}\frac{\delta \hat{%
\mathcal{L}}}{\delta \theta_{~\eta }^{a}}+2\gamma^{bc}\frac{\delta\hat{%
\mathcal{L}}}{\delta \gamma ^{ac}} = 0 ~\text{for}~N \geq 4.
\label{21}
\end{equation}

\subsection{Elements of the Palatini formulation}

In the Palatini formulation there are just two dynamic independent fields: the space-time metric $g^{\mu \nu }$ and the symmetric affine connection $%
\Gamma _{~\mu \nu }^{\lambda }$. The actions are given by
\begin{equation}
I_{n}\left[ g^{\mu \nu },\Gamma _{~\mu \nu }^{\lambda }\right] =\alpha_{n}\int\limits_{\mathcal{M}}\!\hat{\mathcal{L}}_{n}^{\mathcal{P}}\,\mathrm{d%
}V,  \label{22}
\end{equation}%
with%
\begin{equation}
\hat{\mathcal{L}}_{1}^{\mathcal{P}} = \frac{1}{2}\delta _{\alpha \beta}^{\mu \nu }\hat{g}^{\beta \kappa }B_{~\kappa \mu \nu }^{\alpha }; ~\hat{\mathcal{L}}_{2}^{\mathcal{P}} = \frac{1}{4}\delta _{\alpha \beta\gamma \delta }^{\mu \nu \rho \sigma }\hat{g}^{\beta \kappa }g^{\delta\lambda }B_{~\kappa \mu \nu }^{\alpha }B_{~\lambda \rho \sigma }^{\gamma },
\label{23}
\end{equation}%
where
\begin{equation}
\delta _{\alpha \beta }^{\mu \nu }\equiv \delta _{\alpha }^{\mu }\delta
_{\beta }^{\nu }-\delta _{\beta }^{\mu }\delta _{\alpha }^{\nu },
\label{25}
\end{equation}%
and%
\begin{equation}
\delta _{\alpha \beta \gamma \delta }^{\mu \nu \rho \sigma }\equiv \delta
_{\alpha \beta }^{\mu \nu }\delta _{\gamma \delta }^{\rho \sigma }-\delta
_{\alpha \beta }^{\mu \rho }\delta _{\gamma \delta }^{\nu \sigma }+\delta
_{\alpha \beta }^{\mu \sigma }\delta _{\gamma \delta }^{\nu \rho }+\delta _{\alpha \beta }^{\rho \sigma }\delta _{\gamma \delta }^{\mu \nu}-\delta _{\alpha \beta }^{\nu \sigma }\delta _{\gamma \delta }^{\mu \rho}+\delta _{\alpha \beta }^{\nu \rho }\delta _{\gamma \delta }^{\mu \sigma }.
\label{26}
\end{equation}%
The superscript $\mathcal{P}$ denotes Palatini and may appear as a subscript or in parenthesis for clarity, and may be omitted where the context is clear. The variational derivatives of $\hat{\mathcal{L}}_{1}^{P}$ are given by%
\begin{equation}
\frac{1}{\sqrt{\left\vert g\right\vert }}\frac{\delta \hat{\mathcal{L}}_{1}^{%
\mathcal{P}}}{\delta g^{\mu \nu }} = -\frac{1}{2}g_{\mu \nu }\mathcal{L}%
_{1}^{\mathcal{P}}-B_{~\left( \mu \nu \right) \alpha }^{\alpha };~
\frac{\delta \hat{\mathcal{L}}_{1}^{\mathcal{P}}}{\delta \Gamma _{~\mu \nu
}^{\alpha }} = D_{\kappa }\left( \hat{g}^{\beta \left( \mu \right. }\right)
\delta _{\alpha \beta }^{\left. \nu \right) \kappa }.  \label{27}
\end{equation}%
The variational derivative of $\hat{\mathcal{L}}_{2}^{\mathcal{P}}$ with respect to the metric is given by%
\begin{equation}
\frac{1}{\sqrt{\left\vert g\right\vert }}\frac{\delta \hat{\mathcal{L}}_{2}^{%
\mathcal{P}}}{\delta g^{\rho \sigma }}\equiv \mathcal{L}_{\rho \sigma }=-%
\frac{1}{2}g_{\rho \sigma }\mathcal{L}_{2}^{\mathcal{P}}+\frac{1}{2}\delta
_{\alpha \beta \gamma \delta }^{\kappa \lambda \mu \nu }\delta _{\left( \rho
\right. }^{\beta }\delta _{\left. \sigma \right) }^{\eta }B_{~\eta \kappa
\lambda }^{\alpha }B_{~~\mu \nu }^{\gamma \delta }.  \label{29}
\end{equation}%
Note that the second term on the r.h.s. of eq.~(\ref{29}) can be expressed as the sum of the symmetric and antisymmetric parts of $\mathbf{B}$:%
\begin{equation}
\frac{1}{2} \delta _{\alpha \beta \gamma \delta }^{\kappa \lambda \mu \nu }\delta _{\left( \rho \right. }^{\beta }\delta _{\left. \sigma
\right) }^{\eta }B_{~\eta \kappa \lambda }^{\alpha }B_{~~\mu \nu }^{\gamma \delta }= \frac{1}{2}\delta _{\alpha \beta \gamma \delta }^{\kappa \lambda \mu \nu
}\delta _{\left( \rho \right. }^{\beta }g_{\left. \sigma \right) \eta}B_{~~~\kappa \lambda }^{\left[ \alpha \eta \right] }B_{~~\mu \nu }^{\gamma
\delta }+\frac{1}{2}\delta _{\alpha \beta \gamma \delta }^{\kappa \lambda \mu \nu }\delta _{\left( \rho \right. }^{\beta }g_{\left. \sigma \right)
\eta }B_{~~~\kappa \lambda }^{\left( \alpha \eta \right) }B_{~~\mu \nu}^{\gamma \delta }.  \label{30}
\end{equation}%
This way we can rewrite eq.~(\ref{30}) to draw out the \emph{dynamical}
dependence on the non-metricity, $Q_{\mu }^{~\alpha \beta }\equiv D_{\mu
}g^{\alpha \beta }$, which is contained inside $\mathcal{L}_{\rho \sigma}$.
Since%
\begin{equation}
B_{~~~~\kappa \lambda }^{\left( \alpha \beta \right) }=D_{\left[ \kappa\right. }D_{\left. \lambda \right] }g^{\alpha \beta }\equiv D_{\left[ \kappa\lambda \right] }g^{\alpha \beta }=D_{\left[ \kappa \right. }Q_{\left.\lambda \right] }^{~~\alpha \beta },  \label{31}
\end{equation}
we find
\begin{equation}
\mathcal{L}_{\rho \sigma } = -\frac{1}{2}g_{\rho \sigma }\mathcal{L}_{2}^{\mathcal{P}}+\frac{1}{2}\delta _{\alpha \beta \gamma \delta }^{\kappa\lambda \mu \nu }\delta _{\left( \rho \right. }^{\beta }g_{\left. \sigma\right) \eta }B_{~~~\kappa \lambda }^{\left[ \alpha \eta \right] }B_{~~\mu\nu }^{\gamma \delta } +\frac{1}{2}\delta _{\alpha \beta \gamma \delta }^{\kappa \lambda \mu \nu}\delta _{\left( \rho \right. }^{\beta }g_{\left. \sigma \right) \eta}B_{~~\mu \nu }^{\gamma \delta }D_{\kappa }Q_{\lambda }^{~\alpha \eta }.
\label{32}
\end{equation}
The variational derivative with respect to the affine connection is given by%
\begin{equation}
\frac{\delta \hat{\mathcal{L}}_{2}^{\mathcal{P}}}{\delta \Gamma _{~\rho\sigma }^{\lambda }}\equiv \hat{\mathcal{L}}_{\lambda }^{~\rho \sigma}=D_{\kappa }\left( \hat{g}^{\delta \eta }g^{\beta \left( \rho \right.}\right) \delta _{\lambda \beta \gamma \delta }^{\left. \sigma \right)\kappa \mu \nu }B_{~\eta \mu \nu }^{\gamma }.  \label{33}
\end{equation}%
More explicit expressions of eqs.~(\ref{29}) and (\ref{33}) may be found in our previous work \cite{KL1}. Torsion was eliminated from the r.h.s. of eq.~(\ref{33}). Note that eqs.~(\ref{29}) and (\ref{33}) do not contain derivatives of the curvature tensor, which is a property shared by all Lovelock Lagrangians by virtue of the cyclic null property of the curvature tensor in the absence of torsion, $D_{\left( \rho \right\vert }B_{~\lambda \left\vert \mu \nu \right)}^{\kappa }=0$, where $\left( \rho \mu \nu \right) \equiv \frac{1}{3}\left(\rho \mu \nu +\mu \nu \rho +\nu \rho \mu \right) $ is a cyclic permutation of the indices. In section IV we shall introduce a new family of (parity-preserving) Lagrangians that also shares this property.

We emphasize that in general, $Q_{\mu }^{~\alpha \beta }\neq 0$ and $\nabla_{\mu }\gamma ^{ab}\neq 0$ off-shell in the Palatini and the general linear formulation respectively. The independence of the metric and the connection, and consequently the presence of non-metricity in the bulk are necessary to preserve the first condition, namely that the equations of motion contain no higher than first-order derivatives of the fields. It follows that in $N=4$ dimensions we have a non-vanishing contribution from eq.~(\ref{21})
\begin{equation}
\frac{\delta \hat{\mathcal{L}}_{2}^{\mathcal{P}}}{\delta g^{\rho \sigma}}=
\hat{g}_{\eta \left( \rho \right. }\delta _{\left. \sigma \right) }^{\beta
}\delta _{\alpha \beta \gamma \delta}^{\kappa \lambda \mu \nu }B_{~~\kappa
\lambda }^{\gamma \delta}B_{~~~\mu \nu }^{\left( \alpha \eta \right) }.
\label{34}
\end{equation}%
This leads to the generalized Bach-Lanczos identity \cite{HEHL1}. In the absence of torsion, and with a metric-compatible connection, \emph{i.e.} $Q_{\mu}^{~\alpha \beta }=0$, $\hat{\mathcal{L}}_{1}$ and $\hat{\mathcal{L}}_{2}$ are the usual Einstein-Hilbert Lagrangian density $\hat{\mathcal{L}}_{EH}\left[ g_{\mu \nu }\right] $ and the Gauss-Bonnet term $\hat{\mathcal{L}}_{GB}\left[ g_{\mu \nu }\right] $ respectively. In this case the Riemann tensor $B_{~\lambda \mu \nu }^{\kappa }$ reduces to $R_{~\lambda \mu \nu}^{\kappa }$ with the additional symmetries:
\begin{equation}
R_{\kappa \lambda \mu \nu }=R_{\mu \nu \kappa \lambda }~\text{and}~R_{\left(\kappa \lambda \right) \mu \nu }=0.  \label{35}
\end{equation}
It must be emphasized that this reduction occurs only on the boundary at infinity, \emph{i.e.} $\left. B_{~\lambda \mu \nu }^{\kappa }\right\vert _{\partial{\mathcal{M}}}=R_{~\lambda \mu \nu }^{\kappa }$.

\subsection{Symmetries}

The $GL\left( N,\mathbf{R}\right)$ theory, eqs. (\ref{3})-(\ref{7}), is invariant under three gauge symmetries \cite{JS1}:

\begin{enumerate}
\item Local \textquotedblleft frame choice\textquotedblright\ freedom, parameterised by an arbitrary infinitesimal local matrix of $gl(N,\mathbf{R})$, $\lambda _{~b}^{a}=\lambda _{~b}^{a}\left( x\right) $:
\begin{eqnarray}
\delta _{\lambda }\theta _{~\mu }^{a}=\lambda _{~b}^{a}\theta _{~\mu
}^{b},~\delta _{\lambda }\gamma ^{ab}=2\lambda _{~~c}^{\left( a\right.
}\gamma ^{\left. b\right) c},~\delta _{\lambda }\omega _{\mu ~b}^{~a}=-\nabla _{\mu }\lambda
_{~b}^{a}. \label{36}
\end{eqnarray}

\item Local diffeomorphism, parameterised by an arbitrary infinitesimal
vector field, $\xi^{\rho}=\xi^{\rho}\left( x\right) $:%
\begin{equation}
\delta _{\xi }\theta _{~\mu }^{a}=\xi ^{\nu }\partial _{\nu }\theta _{~\mu}^{a}+\theta _{~\nu }^{a}\partial _{\mu }\xi ^{\nu },~\delta _{\xi }\gamma
^{ab}=\xi ^{\mu }\partial _{\mu }\gamma ^{ab},~\delta _{\xi }\omega _{\mu ~b}^{~a}=\xi ^{\nu }\partial _{\nu }\omega _{\mu
~b}^{~a}+\omega _{\nu ~b}^{~a}\partial _{\mu }\xi ^{\nu }.  \label{39}
\end{equation}

\item Projective symmetry, parameterised by an arbitrary infinitesimal
vector field $\kappa _{\mu }=\kappa _{\mu }\left( x\right) $:%
\begin{equation}
\delta _{\kappa }\theta _{~\mu }^{a}=0\text{,\ }\delta _{\kappa }\gamma
^{ab}=0\text{~and~}\delta _{\kappa }\omega _{\mu ~b}^{~a}=\kappa _{\mu
}\delta _{b}^{a}.  \label{40}
\end{equation}
\end{enumerate}

Equations (\ref{36}) - (\ref{40}) may be expressed succinctly as%
\begin{equation}
\delta _{x}y_{A}=\Lambda _{A\Phi } x^{\Phi }+\Lambda _{~A\Phi }^{\mu
}\partial _{\mu }x^{\Phi }~\text{with}~x^{\Phi }=\left\{ \xi ^{\rho },\kappa
_{\mu },\lambda _{~b}^{a}\right\}~\text{and}~y_{A}=\left\{ \gamma ^{ab},\theta _{~\mu }^{a},\omega _{\mu ~b}^{~a}\right\} ,
\label{41}
\end{equation}%
where $y_A$ enumerate collectively the dynamic fields and their respective indices, while $x^{\Phi }$ stands for the symmetry parameters, and capital Greek letters denote their respective indices. Observe that the variations (\ref{41}) contain only first-order derivatives of $y_{A}$'s in $\mathbf{\Lambda }$'s.

The Palatini formulation, eqs.~(\ref{22}) - (\ref{26}), is invariant under local diffeomorphism, in which case%
\begin{eqnarray}
\pounds_{\xi }g^{\mu \nu } & = & \xi ^{\rho }\partial _{\rho }g^{\mu \nu}-2g^{\rho \left( \mu \right. }\partial _{\rho }\xi ^{\left. \nu \right) };
\label{42} \\[1ex]
\pounds _{\xi }\Gamma _{~\mu \nu }^{\lambda } & = &\xi ^{\rho }\partial _{\rho }\Gamma _{~\mu \nu }^{\lambda }+\left( 2\delta _{\left( \mu \right. }^{\rho }\Gamma_{~\left. \nu \right) \kappa }^{\lambda }-\delta _{\kappa }^{\lambda }\Gamma_{~\mu \nu }^{\rho }\right) \partial _{\rho }\xi ^{\kappa } +\partial _{\mu }\partial_{\nu }\xi ^{\lambda },
\label{43}
\end{eqnarray}%
where $\pounds _{\xi }$ is the Lie derivative operator along $\xi ^{\mu }$. In a similar vein to eq.~(\ref{41}), we express eqs.~(\ref{42}) - (\ref{43})
collectively as
\begin{eqnarray}
\pounds _{\xi }y_{A}= \Lambda _{A\rho }\xi ^{\rho }+\Lambda _{A\rho }^{\mu}\partial _{\mu }\xi ^{\rho }+\Lambda _{A\rho }^{\mu \nu }\partial _{\mu \nu
}\xi ^{\rho }~\text{with}~y_{A}=\left\{ g^{\mu \nu },\Gamma _{~\mu \nu }^{\lambda}\right\} .
\label{44}
\end{eqnarray}
Note that the $\mathbf{\Lambda }$'s contain the fields and their first-order derivatives, but that a second-order derivative of $\xi ^{\rho }$ appears in
eq.~(\ref{44}), which does not appear in eq.~(\ref{41}).

\subsection{Irreversibility of Transformations and Field Equations}

The $GL(N,\mathbf{R})$ and Palatini formulations are not interchangeable. One cannot pass back and forth from a space-time, described by the trio $\{\gamma ^{ab},\theta _{~\mu }^{a},\omega _{\mu ~b}^{~a}\}$ to a space-time described by the duo $\{g^{\mu \nu },\Gamma _{~\mu \nu}^{\lambda }\}$, because the transformation is irreversible. The $GL(N,\mathbf{R})$ formulation is endowed with greater structure and symmetry, and in order to go from it to the Palatini formulation one has to break the symmetry by making the canonical (metric) choice on the non-coordinate base:
\begin{equation}
\left. \theta _{~\mu }^{a}\right\vert _{\partial \mathcal{M}}=\delta _{~\mu
}^{a}\Rightarrow \left( \delta _{\xi }+\delta _{\lambda }\right) \theta
_{~\mu }^{a}=0.  \label{45}
\end{equation}
This choice is made only at the boundary, $\partial \mathcal{M}$. This gauge freedom is preserved \cite{JS1} by fixing the residual symmetry by the choice
\begin{equation}
\lambda _{~b}^{a}=-\theta _{b}^{~\mu }\xi ^{\nu }\partial _{\nu }\theta
_{~\mu }^{a}-\theta _{b}^{~\mu }\theta _{~\nu }^{a}\partial _{\mu }\xi ^{\nu
}\Rightarrow \left. \lambda _{~b}^{a}\right\vert _{\partial \mathcal{M}%
}=-\partial _{b}\xi ^{a}.  \label{46}
\end{equation}
It follows that on the boundary
\begin{equation}
-\lambda _{~b}^{a}+\xi ^{\nu }\omega _{\nu ~b}^{~a}=\theta _{b}^{~\mu
}\theta _{~\nu }^{a}D_{\mu }\xi ^{\nu }=\theta _{b}^{~\mu }\nabla _{\mu }\xi
^{a}.  \label{47}
\end{equation}
Even though these formulations are not identical, in many calculations it is implicitly assumed that a result obtained in the more symmetric $GL(N,\mathbf{R})$ formulation (a more computationally convenient setting for some calculations) can be reduced to describe results obtained in the Palatini formulation once symmetry is broken according to (\ref{45}) and (\ref{46}).

This assumption hinges of the dynamics being equal on-shell. Let us explicitly prove this assertion for the case $n=2$. Firstly, we note that by virtue of the vielbein postulate, eq. (\ref{10}), $\hat{\mathcal{L}}_{pq}=\theta _{p}^{~\rho }\theta _{q}^{~\sigma }\hat{\mathcal{L}}_{\rho \sigma }$, so $\left. \hat{\mathcal{L}}_{\rho \sigma }\right\vert _{\partial \mathcal{M}}=0$ implies $\left. \hat{\mathcal{L}}_{pq}\right\vert _{\partial \mathcal{M}}=0$ and vice versa, without the need to resort to eq. (\ref{45}). Secondly, we observe that while $\left. \hat{\mathcal{L}}_{~a}^{\lambda ~b}\right\vert _{\partial \mathcal{M}}=0$ and $\left. \hat{\mathcal{L}}_{\lambda }^{~\left( \rho \sigma \right) }\right\vert _{\partial \mathcal{M}}=0$ are very different in general, the requirement that on the boundary the connections are metric-compatible trivially satisfies both equations. Thirdly, $\hat{\mathcal{L}}_{e}^{~\eta }$ depends on $\nabla _{\lambda }\hat{\mathcal{L}}_{~a}^{\lambda ~b}$ and $\hat{\mathcal{L}}_{pq}$ by virtue of identity (\ref{21}). Since $\nabla _{\lambda }\hat{\mathcal{L}}%
_{~a}^{\lambda ~b}\propto B_{~~~\mu \nu }^{\left( cd\right) }$, and $\left.
B_{~\lambda \mu \nu }^{\kappa }\right\vert _{\partial {\mathcal{M}}%
}=R_{~\lambda \mu \nu }^{\kappa }$, it follows that $\left. \nabla _{\lambda
}\hat{\mathcal{L}}_{~a}^{\lambda ~b}\right\vert _{\partial \mathcal{M}}=0$,
from which $\left. \hat{\mathcal{L}}_{e}^{~\eta }\right\vert _{\partial
\mathcal{M}}=0$ provided that $\left. \hat{\mathcal{L}}_{pq}\right\vert
_{\partial \mathcal{M}}=0$. This implies that with the requirement that $%
\left. \nabla _{\lambda }\gamma ^{ab}\right\vert _{\partial {\mathcal{M}}}=0$%
, $\left. \hat{\mathcal{L}}_{e}^{~\eta }\right\vert _{\partial \mathcal{M}%
}=0 $ does not contain new dynamics. More concretely, $\left. \mathcal{L}
_{e}^{~\eta }\right\vert _{\partial \mathcal{M}}=0$ in eq. (\ref{18}) may be
rewritten as%
\begin{equation}
\theta _{e}^{~\eta }\mathcal{L}_{2}-\theta _{e}^{~\kappa
}\theta_{abcd}^{\eta \lambda \mu \nu }B_{~~\kappa \lambda }^{ab}B_{~~\mu
\nu}^{cd}=0.  \label{48}
\end{equation}%
Multiplying both sides by $\theta _{~\eta }^{f}$ and using eq. (\ref{1}), we
have%
\begin{equation}
\gamma _{ef}\mathcal{L}_{2}-\theta _{e}^{~\kappa }\gamma _{fh}\theta
_{~\eta}^{h}\theta _{abcd}^{\eta \lambda \mu \nu }B_{~~\kappa \lambda
}^{ab}B_{~~\mu \nu }^{cd}=0.  \label{49}
\end{equation}%
This may be split into the symmetric and antisymmetric parts%
\begin{eqnarray}
\gamma _{ef}\mathcal{L}_{2}-\theta _{\left( e\right. }^{~\kappa
}\gamma_{\left. f\right) h}\theta _{~\eta }^{h}\theta _{abcd}^{\eta \lambda
\mu \nu }B_{~~\kappa \lambda }^{ab}B_{~~\mu \nu }^{cd}& =0;  \label{50} \\%
[1ex]
\theta _{\left[ e\right. }^{~\kappa }\gamma _{\left. f\right]
h}\theta_{~\eta }^{h}\theta _{abcd}^{\eta \lambda \mu \nu }B_{~~\kappa
\lambda }^{ab}B_{~~\mu \nu }^{cd}& =0.  \label{51}
\end{eqnarray}%
Since all calculations are done on the boundary, we fix $\theta _{~\eta}^{e}
$ according to eq. (\ref{45}), and take $\left. B_{~\lambda \mu \nu
}^{\kappa }\right\vert _{\partial {\mathcal{M}}}=R_{~\lambda \mu
\nu}^{\kappa }$ with the symmetries in eq. (\ref{35}). We have
\begin{eqnarray}
g_{\rho \sigma }\mathcal{L}_{2}^{\mathcal{P}}-\delta _{\left( \rho
\right.}^{\kappa }g_{\left. \sigma \right) \eta }\delta _{\alpha \beta
\gamma \delta }^{\eta \lambda \mu \nu }R_{~~\kappa \lambda }^{\alpha
\beta}R_{~~\mu \nu }^{\gamma \delta }& =0;  \label{52} \\[1ex]
\delta _{\left[ \rho \right. }^{\kappa }g_{\left. \sigma \right] \eta}\delta
_{\alpha \beta \gamma \delta }^{\eta \lambda \mu \nu }R_{~~\kappa \lambda
}^{\alpha \beta }R_{~~\mu \nu }^{\gamma \delta }& =0.  \label{53}
\end{eqnarray}
The first equation is just $\left. \hat{\mathcal{L}}_{\rho
\sigma}\right\vert _{\partial \mathcal{M}}=0$ in disguise. The second is
identically zero by virtue of $\left. D_{\lambda }g^{\mu \nu
}\right\vert_{\partial {\mathcal{M}}}=0$.~Q.E.D.

We shall show in the next section that the fact that both formulations have
equivalent dynamics is not sufficient for a unique determination of the
superpotential for Lovelock Lagrangians of higher orders, despite their
unique part in the Palatini - metric correspondence \cite{ESJ}.

\section{\label{sec:level3}Superpotentials via Variational Derivatives}

In this section we present the main principles behind and the equations for the variation of the superpotential for the metric-affine theories. A more involved and detailed derivation was given in previous works \cite{S1,JS2,JS1,KL1,KL2}.

\subsection{The superpotential in the general linear formulation}

We begin with the following differential identity:
\begin{equation}
\delta \hat{\mathcal{L}}=\hat{\mathcal{L}}^{A}\delta y_{A}+\partial _{\rho
}\left( \frac{\partial \hat{\mathcal{L}}}{\partial \left( \partial _{\rho
}y_{A}\right) }\delta y_{A}\right) ~\text{with}~\hat{\mathcal{L}}^{A}\equiv
\frac{\delta \hat{\mathcal{L}}}{\delta y_{A}}.  \label{54}
\end{equation}%
Eq.~(\ref{54}) defines the variation of the Lagrangian density due to a continuous variation of the fields $\delta y_{A}$ and their first-order
derivatives. In the case of eq.~(\ref{41}), eq.~(\ref{54}) may be rewritten as the Noether identity:%
\begin{equation}
\hat{\mathcal{L}}^{A}\delta _{X}y_{A}=\partial _{\rho }\left( \xi ^{\rho }
\hat{\mathcal{L}}-\frac{\partial \hat{\mathcal{L}}}{\partial \left( \partial
_{\rho }y_{A}\right) }\delta _{X}y_{A}\right) ,  \label{55}
\end{equation}%
and%
\begin{equation}
\delta _{X}y_{A}\equiv \delta _{\lambda }y_{A}+\delta _{\kappa }y_{A}+\delta
_{\xi }y_{A}.  \label{56}
\end{equation}%
Eq.~(\ref{55}) relates variational derivatives on the left hand-side with other functional derivatives inside a divergence on the right hand-side.
Substituting eq.~(\ref{41}) into both sides of eq.~(\ref{55}), and taking $x^{\Phi }\rightarrow x_{0}^{\Phi }\epsilon $, we can group both sides by the
order of partial derivatives of $\epsilon $ as follows:
\begin{equation}
\epsilon \hat{X}+\left( \partial _{\rho }\epsilon \right) \hat{W}^{\rho}=\epsilon \left( \partial _{\rho }\hat{J}^{\rho }\right) +\left( \partial
_{\rho }\epsilon \right) \left[ \hat{J}^{\rho }+\partial _{\sigma }\hat{U}^{\sigma \rho }\right]+\left( \partial _{\rho \sigma }\epsilon \right) \hat{U}%
^{\left( \rho \sigma \right)},
\label{57}
\end{equation}%
where
\begin{equation}
\hat{X}=\hat{\mathcal{L}}^{A}\Lambda _{A\Phi }x^{\Phi }+\hat{\mathcal{L}}^{A}\Lambda _{A\Phi }^{\rho }\partial _{\rho }x^{\Phi }~\text{and}~\hat{W}%
^{\rho }=\hat{\mathcal{L}}^{A}\Lambda _{A\Phi }^{\rho }x^{\Phi }.
\label{58}
\end{equation}%
$\hat{J}^{\rho }$ and $\hat{U}^{\rho \sigma}$ are functionals of $\mathcal{L}$ and its partial derivatives, and the index $0$ has been removed, since
this identity is now valid in general. Comparing both sides of eq.~(\ref{57}), we obtain three identities:
\begin{equation}
\hat{X}=\partial_{\rho }\hat{J}^{\rho },~\hat{J}^{\rho }+\partial _{\sigma }\hat{U}^{\sigma \rho }=\hat{W}^{\rho }~\text{and}~\hat{U}^{\left( \rho
\sigma \right) }=0.
\label{59}
\end{equation}%
These are known as the Klein identities \cite{PL}. On the boundary, the field equations are satisfied, $\left. \hat{\mathcal{L}}^{A}\right\vert _{\partial \mathcal{M}}=0$, from which follow $\left.
\partial _{\rho }\hat{J}^{\rho }\right\vert _{\partial \mathcal{M}}=0$ and $\left. \hat{W}^{\rho }\right\vert _{\partial \mathcal{M}}=0$. The former
property suggests that $\hat{J}^{\rho }$ is the Noether current density, and since the symmetric part of $\hat{U}^{\rho \sigma }$ vanishes identically,
it follows that $\left. \hat{J}^{\rho }\right\vert _{\partial \mathcal{M}}=\partial _{\sigma }\hat{U}^{\left[ \rho \sigma \right] }$, from which we
deduce that $\hat{U}^{\rho \sigma }$ is the flux or \emph{superpotential}. In general, from the second and third identities in eq.~(\ref{59}) we have:
\begin{equation}
\hat{J}^{\rho }=\hat{W}^{\rho }+\partial _{\sigma }\hat{U}^{\left[\rho \sigma \right] }.
\label{60}
\end{equation}%
In this form the current is a linear sum of the variational derivatives in $\hat{W}^{\rho }$ according to the second equation in (\ref{58}). Since these
variational derivatives contain no higher than first-order derivatives of the fields, a variation of eq.~(\ref{60}) yields:%
\begin{equation}
\delta \hat{J}^{\rho }=\delta \hat{W}^{\rho }+\partial _{\sigma }\delta \hat{U}^{\rho \sigma }=\frac{\delta \hat{W}^{\rho }}{\delta y_{A}}+\partial
_{\sigma }\left( \frac{\partial \hat{W}^{\rho }}{\partial \left( \partial_{\sigma }y_{A}\right) }\delta y_{A}+\delta \hat{U}^{\rho \sigma }\right) .
\label{61}
\end{equation}%
Furthermore, since $\hat{X}$ is a divergence, it follows that (see \cite{KL1}, eq.~($2.23$) for more details)%
\begin{equation}
\frac{\delta \hat{X}}{\delta y_{A}}=0\Rightarrow \frac{\partial \hat{W}^{\left( \rho \right. }}{\partial \left( \partial _{\left. \sigma \right)
}y_{A}\right) }\delta y_{A}=0.
\label{62}
\end{equation}%
Hence, inside the integral, eq.~(\ref{61}) reads:%
\begin{eqnarray}
\delta \left( \int\limits_{\mathcal{M}}\hat{J}^{\rho }dV_{\rho }\right) = \int\limits_{\mathcal{M}}\frac{\delta \hat{W}^{\rho }}{\delta y_{A}}\delta
y_{A}dV_{\rho } +\int\limits_{\partial \mathcal{M}}\left( \frac{\partial \hat{W}^{\left[\rho \right. }}{\partial \left( \partial _{\left. \sigma \right]
}y_{A}\right) }\delta y_{A}+\delta \hat{U}^{\rho \sigma }\right) dS_{\rho\sigma },
\label{63}
\end{eqnarray}%
where $dV_{\rho }$ is the volume element and $dS_{\rho \sigma }$ is the coordinate surface element of a closed surface $S$. The variation of the superpotential
is obtained from the guiding principle, that the boundary term in eq.~(\ref{63}) vanish, so that
\begin{equation}
\left. \delta \hat{U}^{\rho \sigma }\right\vert _{\partial \mathcal{M}}=-\frac{\partial \hat{W}^{\left[ \rho \right. }}{\partial \left( \partial
_{\left. \sigma \right] }y_{A}\right) }\delta y_{A}~\text{with}~y_{A}=\left\{ \gamma ^{ab},\theta _{~\mu }^{a},\omega _{\mu~b}^{~a}\right\} ,
\label{64}
\end{equation}%
and%
\begin{equation}
\hat{W}^{\rho }=\xi ^{a}\frac{\delta \hat{\mathcal{L}}}{\delta \theta_{~\rho }^{a}}+\left( -\lambda _{~b}^{a}+\omega _{\mu ~b}^{~a}\xi ^{\mu
}\right) \frac{\delta \hat{\mathcal{L}}}{\delta \omega _{\rho ~b}^{~a}}.
\label{65}
\end{equation}%
Eqs.~(\ref{64}) and (\ref{65}) appeared originally in \cite{JS1}. The conserved charges that correspond to the symmetries in eqs.~(\ref{36})-(\ref%
{40}) are given by the surface integral:
\begin{equation}
C=\int\limits_{S\left( V\right) }\hat{U}^{\rho \sigma }dS_{\rho \sigma }.
\label{66}
\end{equation}%
Eq.~(\ref{64}) is evaluated on the boundary and must be supplemented with boundary conditions. These are%
\begin{equation}
\left. \theta _{~\mu }^{a}\right\vert _{\partial \mathcal{M}}=\delta _{~\mu}^{a},~\left. \left(\nabla _{\mu }\gamma ^{ab}\right)\right\vert _{\partial \mathcal{M}%
}=0~\text{and}~\left. \delta \gamma ^{ab}\right\vert _{\partial \mathcal{M}}=0.
\label{67}
\end{equation}%
The first condition in eq.~(\ref{67}) breaks the symmetry according to eq.~(\ref{45}), thus recreating metric GR on the boundary, with vanishing
torsion and non-metricity (the second condition). The third condition in eq.~(\ref{67}) imposes Dirichlet boundary conditions on the fibre metric.

Moreover, in the case of Lovelock Lagrangians, eqs.~(\ref{3})-(\ref{7}), the second term in the r.h.s. of eq.~(\ref{65}) cannot contribute to the
superpotential on the boundary, since the variational derivative with respect to the spin connection, $\hat{\mathcal{L}}_{~a}^{\rho ~b}$, eqs.~(\ref{17}) and (\ref{20}), differentiated with respect to the partial derivative of the spin connection, will produce terms which are proportional to the non-metricity.
Taking all this into account leaves only the first term in the r.h.s. of eq.~(\ref{65}), which contributes to the sum in eq.~(\ref{64}):
\begin{equation}
\left. \delta \hat{U}^{\rho \sigma }\right\vert _{\partial \mathcal{M}}=-\frac{\partial \hat{W}^{\left[ \rho \right. }}{\partial \left( \partial
_{\left. \sigma \right] }\omega _{\mu ~b}^{~a}\right) }\delta \omega _{\mu~b}^{~a}~\text{with}~\hat{W}^{\rho }\overset{\bullet }{=}\xi ^{a}\frac{%
\delta \hat{\mathcal{L}}}{\delta \theta _{~\rho }^{a}},
\label{68}
\end{equation}%
where we have defined a useful symbol, $A\overset{\bullet }{=}B$, which denotes the relevant part of $A$ is $B$. It is eq.~(\ref{68}) that we shall
use in this paper. In the case $n=1$, eq.~(\ref{4}), differentiating eq.~(\ref{15}) with respect to $\partial _{\sigma }\omega _{\mu ~b}^{~a}$, and
taking the antisymmetric part in $\rho $ and $\sigma $, we find
\begin{equation}
\left. \delta \hat{U}_{1}^{\rho \sigma }\left[ g,\Gamma \right] \right\vert
_{\partial \mathcal{M}}=-3\alpha _{1}\xi ^{\left( \rho \right. }\delta _{\mu
\nu }^{\left. \sigma \lambda \right) }\hat{g}^{\nu \kappa }\delta \Gamma
_{\kappa \lambda }^{\mu },  \label{69}
\end{equation}%
where we have used eq.~(\ref{10}). Since the superpotential is evaluated at
the boundary, the factors of $\delta \Gamma _{~\kappa \lambda }^{\mu }$ are
defined entirely on the boundary \cite{KL2}, so that we may integrate to
obtain%
\begin{equation}
\frac{1}{\alpha _{1}}\hat{U}_{1}^{\rho \sigma }=-3\bar{\xi}^{\left( \rho
\right. }\delta _{\mu \nu }^{\left. \sigma \lambda \right) }\bar{\hat{g}}%
^{\nu \kappa }\Delta _{\kappa \lambda }^{\mu },  \label{70}
\end{equation}%
where, we have introduced the background metric \cite{KBL} $\bar{g}^{\mu \nu }$ , and the background affine connection $\bar{\Gamma}_{~\mu \nu}^{\rho}$ through
\begin{equation}
\Delta _{~\mu \nu }^{\rho }\equiv \Gamma _{~\mu \nu }^{\rho }-\bar{\Gamma}_{~\mu \nu }^{\rho }.
\label{71}
\end{equation}%
For $\alpha _{1}=c^{4}/16\pi G$, eq. (\ref{70}) is the KBL superpotential in disguise \cite{KL1}. Similarly, for $n=2$ we find
\begin{equation}
\frac{1}{\alpha _{2}}\hat{U}_{2}^{\rho \sigma }=-\bar{\xi}^{\eta }\delta_{\alpha \beta \gamma \delta \eta }^{\rho \sigma \mu \nu \kappa }\bar{\hat{g}%
}^{\beta \lambda }R_{~~\mu \nu }^{\gamma \delta }\Delta _{~\kappa \lambda}^{\alpha }.
\label{72}
\end{equation}%
$\hat{U}_{2}^{\rho \sigma }$ possesses several remarkable properties \cite{KL2}. In particular, it vanishes identically in $N=4$ dimensions, which is consistent with the metric formulation, in which case $\hat{\mathcal{L}}_{2}^{\mathcal{P}}$ is a pure divergence in $4$ dimensions.

To see this explicitly, note that $R_{~\delta \mu \nu }^{\gamma }$ in eq.~(\ref{72}) is the Riemann curvature tensor of the background space-time, obeying the symmetries in eq.~(\ref{35}), and as such it may be decomposed into the irreducible representation, with the associated Weyl tensor, $\bar{C}_{\alpha \beta \gamma \delta }$, trace-free Ricci tensor, $r_{\alpha \beta }\equiv \bar{R}_{\alpha \beta }-\frac{1}{N}\bar{g}_{\alpha \beta }\bar{R}$, and scalar curvature, $\bar{R}$, all of which are background quantities. Thus
\begin{equation}
\begin{split}
R_{\mu \nu \rho \sigma } & = \bar{C}_{\mu \nu \rho \sigma }+\frac{2}{N-2}\left( \bar{g}_{\mu \left[ \rho \right. }r_{\left. \sigma \right] \nu }-\bar{%
g}_{\nu \left[ \rho \right. }r_{\left. \sigma \right] \mu }\right) +\frac{2}{N\left( N-1\right) }\bar{g}_{\mu \left[ \rho \right. }\bar{g}%
_{\left. \sigma \right] \nu }\bar{R} \\
& \equiv  C_{\mu \nu \rho \sigma }+E_{\mu \nu \rho \sigma }+G_{\mu \nu \rho \sigma }.
\end{split}
\label{73}
\end{equation}%
Substituting eq.~(\ref{73}) into the r.h.s. of eq.~(\ref{72}), we find%
\begin{equation}
\begin{split}
\frac{1}{\alpha _{2}}U_{2}^{\rho \sigma }=& ~-9C_{~~~\left( \kappa \lambda
\right. }^{\left( \rho \sigma \right. }\delta _{\left. \nu \right) }^{\left.
\mu \right) }\bar{g}^{\nu \eta }\xi ^{\kappa }\Delta _{~\mu \eta }^{\lambda }
-\frac{9}{2}\left( N-4\right) E_{~~~\left( \kappa \lambda \right. }^{\left( \rho \sigma \right. }\delta _{\left. \nu \right) }^{\left. \mu \right)}\bar{g}^{\nu \eta }\bar{\xi}^{\kappa }\Delta _{~\mu \eta }^{\lambda } \\
& ~+\frac{3}{2}\left( N-3\right) \left( N-4\right) \xi ^{\left( \mu \right.
}G_{~~~~\lambda }^{\left. \rho \sigma \right) \nu }\Delta _{~\mu \nu
}^{\lambda }.
\end{split}
\label{74}
\end{equation}
The first term is r.h.s. of eq.~(\ref{74}) is the Bach-Lanczos identity, which vanishes in four dimensions. If space-time is asymptotically maximally
symmetric, \emph{i.e.} $C_{\mu \nu \rho \sigma }=0$ and $r_{\alpha \beta }=0$, then eq.~(\ref{74}) simplifies considerably \cite{KL1,KL2} to
\begin{equation}
\hat{U}_{2}^{\rho \sigma} = \frac{3\alpha_{2}}{2}\left( N-3\right) \left( N-4\right) \bar{\hat{\xi}}^{\left( \mu \right. }G_{~~~~~\lambda
}^{\left. \rho \sigma \right) \nu }\Delta _{~\mu \nu }^{\lambda} = 2\frac{\alpha_{2}}{\alpha_{1}}\frac{\left(N-3\right)\left(N-4\right)}{N\left(N-1\right)}\bar{R}\hat{U}^{\rho \sigma}_{1}.
\label{75}
\end{equation}

\subsection{The superpotential in the Palatini formulation}

In the Palatini formulation, the only relevant symmetry is diffeomorphism invariance. Similarly to eq.~(\ref{55}), we have the following differential identity:%
\begin{equation}
\hat{\mathcal{L}}_{\mathcal{P}}^{A}\pounds _{\xi }y_{A}\equiv \frac{\delta \hat{\mathcal{L}}_{\mathcal{P}}}{\delta y_{A}}\pounds _{\xi }y_{A}=\partial_{\rho }\left( \hat{\mathcal{L}}_{P}\xi ^{\rho }-\frac{\partial \hat{\mathcal{L}}_{\mathcal{P}}}{\partial \left( \partial _{\rho }y_{A}\right) }\pounds _{\xi }y_{A}\right) .  \label{76}
\end{equation}%
We now repeat the same procedure as in eq.~(\ref{57}) with some qualifications. Substituting eq.~(\ref{44}) into both sides of eq.~(\ref{76}), and taking $\xi ^{\lambda }\rightarrow \xi _{0}^{\lambda }\epsilon $, we can group both sides by the order of partial derivatives of $\epsilon$ as follows:
\begin{equation}
\begin{split}
\epsilon \hat{X}_{\mathcal{P}}+\left( \partial _{\rho }\epsilon \right)\hat{W}_{\mathcal{P}}^{\rho }+\left( \partial _{\rho \sigma }\epsilon\right) \hat{Y}^{\rho \sigma } = &~\epsilon \left( \partial _{\rho }\hat{J}_{\mathcal{P}}^{\rho }\right)+\left( \partial _{\rho }\epsilon \right) \left[ \hat{J}_{\mathcal{P}}^{\rho}+\partial _{\sigma }\hat{u}^{\sigma \rho }\right] \\
& +\left( \partial _{\rho \sigma }\epsilon \right) \left[ \hat{u}^{\left( \rho \sigma \right) }+\partial _{\lambda }\hat{V}^{\lambda \left(\rho \sigma \right) }\right] +\left( \partial _{\eta \rho \sigma }\epsilon\right) \hat{V}^{\eta \rho \sigma },
\end{split}
\label{77}
\end{equation}%
where%
\begin{eqnarray}
\hat{X}_{\mathcal{P}} & \equiv & \hat{\mathcal{L}}_{\mathcal{P}}^{A}\Lambda_{A\lambda }\xi ^{\lambda }+\hat{\mathcal{L}}_{\mathcal{P}}^{A}\Lambda
_{A\lambda }^{\rho }\partial _{\rho }\xi ^{\lambda }+\hat{\mathcal{L}}_{\mathcal{P}}^{A}\Lambda _{A\lambda }^{\left( \rho \sigma \right) }\partial_{\rho \sigma }\xi ^{\lambda };  \label{78} \\[0.01in]
\hat{W}_{\mathcal{P}}^{\rho } & \equiv & \hat{\mathcal{L}}_{\mathcal{P}}^{A}\Lambda _{A\lambda }^{\rho }\xi ^{\lambda }+2\hat{\mathcal{L}}_{\mathcal{P}}^{A}\Lambda _{A\lambda }^{\left( \rho \sigma \right) }\partial_{\sigma }\xi ^{\lambda };  \label{79} \\[0.01in]
\hat{Y}^{\rho \sigma } & \equiv & \hat{\mathcal{L}}_{\mathcal{P}}^{A}\Lambda_{A\lambda }^{\left( \rho \sigma \right)}\xi ^{\lambda },
\label{80}
\end{eqnarray}%
and $\hat{J}_{\mathcal{P}}^{\rho }$, $\hat{u}^{\rho \sigma }$ and $\hat{V}^{\eta \rho \sigma }$ are functions of $\hat{\mathcal{L}}_{\mathcal{P}}$ and its functional derivatives. Eqs.~(\ref{77})-(\ref{80}) are more complicated than eqs.~(\ref{57})-(\ref{58}) owing to the second-order derivatives that appear in eq.~(\ref{44}). The entire derivation is given in \cite{KL1}. To make the treatment self-contained, we quote here the main results. Comparing both sides of eq.~(\ref{77}), we obtain four identities:
\begin{equation}
\partial _{\rho }\hat{J}_{\mathcal{P}}^{\rho } = \hat{X}_{\mathcal{P}},\hat{W}_{\mathcal{P}}^{\rho }=\hat{J}_{\mathcal{P}}^{\rho }+\partial _{\sigma } \hat{u}^{\sigma \rho }, \hat{Y}^{\rho \sigma }=\hat{u}^{\left( \rho \sigma \right) }+\partial_{\lambda }\hat{V}^{\lambda \left( \rho \sigma \right) }~\text{and}~\hat{V}^{\left( \lambda \rho \sigma \right) }=0.
\label{81}
\end{equation}%
Solving the identities in eq.~(\ref{81}), we find an expression for the current density, $\hat{J}_{\mathcal{P}}^{\rho }$, in terms of variational derivatives and their partial derivatives:%
\begin{equation}
\hat{J}_{\mathcal{P}}^{\rho }=\hat{W}_{\mathcal{P}}^{\rho }-\partial_{\sigma }\hat{Y}^{\rho \sigma }+\partial _{\sigma }\hat{U}_{\mathcal{P}}^{\rho \sigma }.
\label{82}
\end{equation}%
Most substantially, the current manifests explicit dependence on second-order derivatives of the fields that are contained in the second term, $\partial_{\sigma }\hat{Y}^{\rho \sigma}$. This fact implies that the boundary conditions, eq.~(\ref{67}), that were used to solve eq.~(\ref{64}) in the general linear formulation are insufficient for the solution of the Palatini problem. The variation of the current yields:
\begin{equation}
\delta \hat{J}_{\mathcal{P}}^{\rho }=\delta \hat{W}_{\mathcal{P}}^{\rho}-\partial _{\sigma }\delta \hat{Y}^{\rho \sigma }+\partial _{\sigma }\delta
\hat{U}_{\mathcal{P}}^{\rho \sigma }.
\label{83}
\end{equation}%
The variational derivatives of $\hat{W}_{\mathcal{P}}^{\rho }$ and $\hat{Y}^{\rho \sigma }$ are calculated, and then expressed via the identities in
terms of $\hat{u}^{\rho \sigma }$ and $\hat{V}^{\lambda \rho \sigma }$. Use is made of the following identity (see \cite{KL1}, eq.~($3.20$))%
\begin{equation}
\partial _{\sigma \lambda }\hat{Y}^{\lambda \rho \sigma }=-\frac{2}{3}%
\partial _{\sigma }\left( \partial _{\lambda }\hat{Y}^{\left[ \rho \sigma %
\right] \lambda }\right) ~\text{with}~\hat{Y}^{\lambda \rho \sigma }\equiv
\frac{\partial \hat{Y}^{\rho \sigma }}{\partial \left( \partial _{\lambda
}y_{A}\right) }\delta y_{A}.  \label{84}
\end{equation}%
Following the same guiding principle, we find an equation for the variation of the superpotential $\delta \hat{U}_{\mathcal{P}}^{\rho \sigma }$ by
demanding that the boundary terms vanish. The equation reads:%
\begin{equation}
\left. \delta \hat{U}_{\mathcal{P}}^{\rho \sigma }\right\vert _{\partial\mathcal{M}} =-\frac{\partial \hat{W}_{\mathcal{P}}^{\left[ \rho \right. }}{\partial
\left( \partial _{\left. \sigma \right] }y_{A}\right) }\delta y_{A}+\frac{2}{3}\partial _{\lambda }\left( \frac{\partial \hat{Y}^{\lambda \left[ \rho
\right. }}{\partial \left( \partial _{\left. \sigma \right] }y_{A}\right) }\delta y_{A}\right)~\text{with}~y_{A}=\left\{ g^{\mu \nu },\Gamma _{~\mu \nu }^{\lambda
}\right\}.
\label{85}
\end{equation}%
Here the boundary conditions necessitate not only the vanishing of non-metricity at the boundary, but also of its variation. The latter is consistent with our requirement for metric-compatibility (see Appendix for details). We have%
\begin{equation}
\left. Q_{\lambda }^{~\mu \nu }\right\vert _{\partial \mathcal{M}}=0,~\left.
\delta Q_{\lambda }^{~\mu \nu }\right\vert _{\partial \mathcal{M}}=0~\text{%
and}~\left. \delta g^{\mu \nu }\right\vert _{\partial \mathcal{M}}=0.
\label{86}
\end{equation}%
In particular, since%
\begin{equation}
\partial _{\lambda }\delta g^{\mu \nu }=\delta Q_{\lambda }^{~\mu \nu
}-2\delta g^{\kappa (\mu }\Gamma _{~\kappa \lambda }^{\nu )}-2g^{\kappa (\mu
}\delta \Gamma _{~\kappa \lambda }^{\nu )},  \label{87}
\end{equation}%
it follows from eq.~(\ref{86}) that on the boundary%
\begin{equation}
\left. \left( \partial _{\lambda }\delta g^{\mu \nu }\right) \right\vert
_{\partial \mathcal{M}}=-2g^{\kappa (\mu }\delta \Delta _{~\kappa \lambda
}^{\nu )}.  \label{88}
\end{equation}%
Taking into account the functional dependence of the variational derivative with respect to $\Gamma^{\lambda}_{~\mu \nu}$ on $Q_{\lambda}^{~\mu \nu}$ (see, e.g., eqs.~(\ref{27}) and (\ref{33})), and the boundary conditions in eq.~(\ref{86}), after integration the superpotential becomes:%
\begin{equation}
\left. \hat{U}_{\mathcal{P}}^{\rho \sigma }\right\vert _{\partial \mathcal{M}
}=-\frac{\partial \hat{W}_{\mathcal{P}}^{\left[ \rho \right. }}{\partial \left( \partial
_{\left. \sigma \right] }\Gamma _{~\mu \nu }^{\lambda }\right) }\Delta
_{~\mu \nu }^{\lambda }-\frac{4}{3}\frac{\partial \hat{Y}^{\lambda \left[
\rho \right. }}{\partial \left( \partial _{\left. \sigma \right] }g^{\mu \nu
}\right) }g^{\kappa \left( \mu \right. }\Delta _{~~\kappa \lambda }^{\left.
\nu \right) },  \label{89}
\end{equation}%
where%
\begin{equation}
\hat{W}_{\mathcal{P}}^{\rho }=-2g^{\rho \mu }\xi ^{\nu }\hat{\mathcal{L}}_{\mathcal{\mu \nu }}+2\hat{\mathcal{L}}_{\lambda }^{~\left( \rho \sigma
\right) }D_{\sigma }\xi ^{\lambda }-\hat{Y}^{\mu \nu }\Gamma _{~\mu \nu}^{\rho }~\text{and}~\hat{Y}^{\rho \sigma }=\xi ^{\lambda }\hat{\mathcal{L}}%
_{\lambda }^{~\left( \rho \sigma \right) }.
\label{90}
\end{equation}%

For $\hat{\mathcal{L}}_{1}^{\mathcal{P}}$ we find the KBL superpotential in complete agreement with eq.~(\ref{70}). For $\hat{\mathcal{L}}_{2}^{\mathcal{P}}$, the second equation in eq.~(\ref{23}), we find:
\begin{equation}
\left. \frac{1}{\alpha _{2}}\hat{U}_{2}^{\rho \sigma }\left( \mathcal{P}\right) \right\vert _{\partial \mathcal{M}} =
-\bar{\xi}^{\eta }\delta_{\alpha \beta \gamma \delta \eta }^{\rho \sigma \mu \nu \kappa} R_{~~\mu\nu }^{\gamma \delta}\bar{\hat{g}}^{\beta \lambda }\Delta _{~\kappa \lambda}^{\alpha} +\frac{2}{3}\bar{\xi}^{\alpha }\delta _{\alpha \beta \gamma \delta
}^{\mu \nu \rho \sigma }R_{~~\lambda \mu }^{\gamma \delta }\bar{\hat{g}}
^{\kappa \left( \lambda \right. }\Delta _{~~\nu \kappa }^{\left. \beta
\right)}= \hat{U}_{2}^{\rho \sigma }+{\Delta \hat{U}}_{2}^{\rho \sigma }.
\label{91}
\end{equation}

$\hat{U}_{2}^{\rho \sigma }\left( \mathcal{P}\right)$ can be expressed as the sum of $33$ terms \cite{KL1}. The first term on the r.h.s. of eq.~(\ref{91}) is the same as $\hat{U}_{2}^{\rho \sigma }$ in eq.~(\ref{72}), which is the superpotential in the general linear formulation. Then we have a novelty: the second term, ${\Delta \hat{U}}_{2}^{\rho \sigma }$, which contains $18$ additional terms, cannot be obtained from eq.~(\ref{68}). Although it might not be obvious, all $18$ terms in ${\Delta \hat{U}}_{2}^{\rho \sigma }$ actually appear in $\hat{U}_{2}^{\rho \sigma }$, but with different numerical factors \cite{KL1}. It is only after these terms are isolated, and grouped into a single sum, that their unique distinguishing symmetry becomes apparent. Namely, $\hat{U}_{2}^{\rho \sigma }$ is characterised by an asymmetry in $\beta $ and $\lambda $ in $\Delta _{~\kappa \lambda}^{\beta }$, whereas ${\Delta \hat{U}}_{2}^{\rho \sigma }$ is manifestly symmetric under the exchange of those two indices. We shall show in section IV that this symmetry is linked to the appearance of terms with explicit non-metricity.

It is quite remarkable that the scalar curvature is not present in ${\Delta U}_{2}^{\rho \sigma }$. Indeed, written in terms of the irreducible
background tensors, it is given by:
\begin{equation}
\begin{split}
{\frac{1}{\alpha _{2}}}{\Delta U}_{2}^{\rho \sigma } = &~2\left( \Delta _{~\mu
\nu }^{\kappa }\bar{g}^{\mu (\nu }C_{~~~\kappa \lambda }^{\rho \sigma
)}+\Delta _{~~\mu \nu }^{(\mu }C_{~~~\kappa \lambda }^{\rho \sigma )}\bar{g}%
^{\kappa \nu }\right) \bar{\xi}^{\lambda } \\
& + 2 \left( \frac{N-3}{N-2}\right) \left[
\begin{array}{c}
r^{\mu (\rho }\Delta _{~\mu \nu }^{\sigma }\bar{\xi}^{\nu )}-r^{\mu (\sigma
}\Delta _{~\mu \nu }^{\rho }\bar{\xi}^{\nu )} \\[1ex]
-\Delta _{~\mu \nu }^{\lambda }\left( \bar{g}^{\mu (\rho }r_{~\lambda
}^{\sigma }\bar{\xi}^{\nu )}-\bar{g}^{\mu (\sigma }r_{~\lambda }^{\rho }\bar{%
\xi}^{\nu )}\right)%
\end{array}
\right] .
\end{split}
\label{92}
\end{equation}

From eq.~(\ref{92}) it is clear that ${\Delta U}_{2}^{\rho \sigma }$ has a
non-zero contribution to the superpotential in $4$ dimensions, but vanishes
identically in $3$ dimensions. Furthermore, it is evident from eq.~(\ref{92}) that in order for both formulations to agree exactly, it must follow either that ${\Delta \hat{U}}_{2}^{\rho \sigma }=0$, or more generally that $\int {\Delta \hat{U}}_{2}^{\rho \sigma }dS_{\rho \sigma }=0$. We suspect that the latter induces the former on the boundary, for ordinary (non-turbulent) flow. In the first case, it follows that $C_{~\nu \rho \sigma }^{\mu }=0$ and $r_{\mu \nu}=0$, which means space-time must be asymptotically maximally symmetric.

In the Appendix we show that the conclusions and the results we have reached for $n=2$ apply to higher orders.

We note here that caution must be exercised when comparing the solution in eq.~(\ref{91}) with the equation for the superpotential in eq.~(\ref{85}). The terms that appear with a factor of $2/3$ in eq.~(\ref{91}) do not correspond to the second term in eq.~(\ref{85}), but are obtained from both terms!

\subsection{The plot thickens: $GL(N,\mathbf{R}) \subseteq$ Palatini}

We have shown in the previous section that the superpotentials obtained from the general linear and Palatini formulations differ for $n=2$, even though they are obtained from a series of identities based on the same guiding principle.

Nevertheless, here we show that it is possible to express the Palatini current density $\hat{J}_{\mathcal{P}}^{\rho }$ in eq.~(\ref{82}) in the case of Lovelock Lagrangians in a way that reproduces the superpotential of the general linear formulation, eq.~(\ref{68})!

To this end, we must break the symmetry of $\hat{\mathcal{L}}_{\lambda}^{~\left( \rho \sigma \right) }$ in eq.~(\ref{33}). The latter is
manifestly symmetric in $\rho $ and $\sigma $, but this symmetry is broken when we apply covariant differentiation with respect to $x^{\sigma }$ as follows:
\begin{equation}
\begin{split}
D_{\sigma }\hat{\mathcal{L}}_{\lambda }^{~\left( \rho \sigma \right) }= & ~
\frac{1}{2}\delta _{\lambda \beta \gamma \delta }^{\rho \mu \eta \tau
}D_{\sigma }\left[ B_{~\kappa \eta \tau }^{\gamma }D_{\mu }\left( \hat{g}%
^{\beta \sigma }g^{\delta \kappa }\right) \right] +\frac{1}{2}\delta _{\lambda \beta \gamma \delta }^{\mu \nu \eta \tau}B_{~\kappa \eta \tau }^{\gamma }D_{\mu \nu }\left( \hat{g}^{\beta \rho}g^{\delta \kappa }\right) \\
\equiv &~\frac{1}{2}D_{\sigma }\hat{\mathcal{L}}_{\lambda }^{~\rho \sigma }+%
\frac{1}{2}D_{\sigma }\hat{\mathcal{L}}_{\lambda }^{~\sigma \rho }.
\end{split}
\label{93}
\end{equation}%
The first term, $\frac{1}{2}D_{\sigma }\hat{\mathcal{L}}_{\lambda }^{~\rho
\sigma }$, contains second-order derivatives of $\Gamma _{~\mu \nu
}^{\lambda }$ and $g^{\mu \nu }$, while the second term, $\frac{1}{2}%
D_{\sigma }\hat{\mathcal{L}}_{\lambda }^{~\sigma \rho }$, contains only
first-order derivatives of the affine connection. Indeed, using eq.~(\ref{31}%
), we have
\begin{equation}
D_{\sigma }\hat{\mathcal{L}}_{\lambda }^{~\sigma \rho } =
\delta _{\lambda \beta \gamma \delta }^{\mu \nu \eta \tau }B_{~\kappa
\eta \tau }^{\gamma }\left( \hat{g}^{\delta \kappa }B_{~~~~\mu \nu }^{\left(
\beta \rho \right) }+\hat{g}^{\beta \rho }B_{~~~~\mu \nu }^{\left( \delta
\kappa \right) }-\frac{1}{2}\hat{g}^{\beta \rho }g^{\delta \kappa
}B_{~\alpha \mu \nu }^{\alpha }\right).
\label{94}
\end{equation}
Let us examine $\hat{J}_{\mathcal{P}}^{\rho }$ in eq.~(\ref{82}) in greater
detail. Substituting $\hat{W}_{\mathcal{P}}^{\rho }$ and $\hat{Y}^{\rho
\sigma }$ from eq.~(\ref{90}) into the r.h.s. of eq.~(\ref{82}), we find
\begin{eqnarray}
\hat{J}_{\mathcal{P}}^{\rho } & = & -2g^{\rho \mu }\xi ^{\nu }\hat{\mathcal{L}}_{%
\mathcal{\mu \nu }}+2\hat{\mathcal{L}}_{\lambda }^{~\left( \rho \sigma
\right) }D_{\sigma }\xi ^{\lambda } - \left( \partial _{\sigma }\hat{Y}^{\rho \sigma }+\hat{Y}^{\mu \nu }\Gamma_{~\mu \nu }^{\rho }\right) +\partial _{\sigma }\hat{U}_{\mathcal{P}}^{\prime \rho \sigma }  \label{95} \\
& = & -2g^{\rho \mu }\xi ^{\nu }\hat{\mathcal{L}}_{\mathcal{\mu \nu }}-2\xi
^{\lambda }D_{\sigma }\hat{\mathcal{L}}_{\lambda }^{~\left( \rho \sigma
\right) }+D_{\sigma }\hat{Y}^{\rho \sigma }+\partial _{\sigma }\hat{U}_{%
\mathcal{P}}^{\prime \rho \sigma }  \nonumber \\
& = & -\xi ^{\nu }\left( 2g^{\rho \mu }\hat{\mathcal{L}}_{\mathcal{\mu \nu }%
}+D_{\sigma }\hat{\mathcal{L}}_{\nu }^{~\sigma \rho }\right) -\xi ^{\lambda
}D_{\sigma }\hat{\mathcal{L}}_{\lambda }^{~\rho \sigma }+D_{\sigma }\hat{Y}%
^{\rho \sigma } + \partial _{\sigma }\hat{U}_{\mathcal{P}}^{\prime \rho \sigma },  \nonumber
\end{eqnarray}%
where a prime has been added to distinguish $\hat{U}_{\mathcal{P}}^{\prime
\rho \sigma }$ from $\hat{U}_{\mathcal{P}}^{\rho \sigma }$. Denote%
\begin{equation}
\hat{\jmath}_{~\nu }^{\rho }\equiv 2g^{\rho \mu }\hat{\mathcal{L}}_{\mu \nu
}+D_{\sigma }\hat{\mathcal{L}}_{\nu }^{~\sigma \rho }.  \label{96}
\end{equation}%
Contrast $\hat{\jmath}_{~\nu }^{\rho }$ in eq.~(\ref{96}) with identity (\ref{21}). It turns out that $\hat{\jmath}_{~\nu }^{\rho }$ is nothing but $-\theta^{a}_{~\nu}\frac{\delta \hat{\mathcal{L}}}{\delta \theta_{~\rho}^{a}}$ after the breaking of symmetry in eqs.~(\ref{45}) and (\ref{47}). From this very term we have derived $\hat{U}^{\rho \sigma}$ in the $\mathbf{GL}(N,R)$ formulation. Moreover, the second and third terms in the last equality in eq.~(\ref{95}) may be rewritten thus
\begin{eqnarray}
-\xi^{\lambda }D_{\sigma} \hat{\mathcal{L}}_{\lambda }^{~\rho \sigma
}+D_{\sigma }\hat{Y}^{\rho \sigma} & = &-\xi ^{\lambda }D_{\sigma }\hat{\mathcal{%
L}}_{\lambda }^{~\rho \sigma }+D_{\sigma }\left( \hat{\mathcal{L}}_{\lambda
}^{~\left( \rho \sigma \right) }\xi ^{\lambda }\right) =-\xi ^{\lambda }D_{\sigma }\hat{\mathcal{L}}_{\lambda }^{~\left[ \rho \sigma \right] }+\hat{\mathcal{L}}_{\lambda }^{~\left( \rho \sigma \right)} D_{\sigma }\xi ^{\lambda }  \label{97} \\
& = & -\partial _{\sigma }\left( \xi ^{\lambda }\hat{\mathcal{L}}_{\lambda }^{~%
\left[ \rho \sigma \right] }\right) +\left( \hat{\mathcal{L}}_{\lambda }^{~%
\left[ \rho \sigma \right] }+\hat{\mathcal{L}}_{\lambda }^{~\left( \rho
\sigma \right) }\right) D_{\sigma }\xi ^{\lambda }.  \nonumber
\end{eqnarray}%
Replace eq.~(\ref{96}) and eq.~(\ref{97}) into the Noether current (\ref{95}%
), and expand:%
\begin{equation}
\hat{J}_{\mathcal{P}}^{\rho }=-\xi ^{\nu }\hat{\jmath}_{~\nu }^{\rho }+\hat{%
\mathcal{L}}_{\lambda }^{~\rho \sigma }D_{\sigma }\xi ^{\lambda }+\partial
_{\sigma }\left( \hat{U}_{\mathcal{P}}^{\prime \rho \sigma }-\xi ^{\lambda }%
\hat{\mathcal{L}}_{\lambda }^{~\left[ \rho \sigma \right] }\right) .
\label{98}
\end{equation}%
$\hat{\mathcal{L}}_{\lambda }^{~\rho \sigma }$ that appears in the second
term is no longer symmetric here (since this symmetry was previously broken
into two parts!), and should not be confused with eq.~(\ref{33}). The first
and second terms in the r.h.s. of eq.~(\ref{98}) are equal to the current in
the $\mathbf{GL}(N,R)$ formulation (after the symmetry breaking and
gauge fixing), and we denote this fact by%
\begin{equation}
\hat{W}_{\text{GL}}^{\rho } \equiv -\xi ^{\nu }\hat{\jmath}_{~\nu }^{\rho }+
\hat{\mathcal{L}}_{\lambda }^{~\rho \sigma }D_{\sigma }\xi ^{\lambda }=-\xi ^{\nu }\left( 2g^{\rho \mu }\hat{\mathcal{L}}_{\mu \nu }+D_{\sigma }\hat{\mathcal{L}}_{\nu }^{~\sigma \rho }\right) +\hat{\mathcal{L}}_{\lambda
}^{~\rho \sigma }D_{\sigma }\xi ^{\lambda }.
\label{99}
\end{equation}%
By virtue of eqs.~(\ref{29}) and (\ref{94}), $\hat{W}_{\text{GL}}^{\rho }$ contains only first-order derivatives of the fields. The boundary $\partial \mathcal{M}$ is defined by the vanishing of the variational derivatives, and together with the vanishing of the non-metricity at the boundary, we find that  $\hat{W}_{\text{GL}}^{\rho }=0$ on-shell. Now denote
\begin{equation}
\begin{split}
\hat{Z}^{\rho \sigma } & \equiv -\xi ^{\lambda }\hat{\mathcal{L}}_{\lambda}^{~\left[ \rho \sigma \right] }=\frac{1}{2}\xi ^{\alpha }\delta _{\alpha\beta \gamma \delta }^{\mu \rho \eta \tau }B_{~\kappa \eta \tau }^{\gamma}D_{\mu }\left( \hat{g}^{\beta \sigma }g^{\delta \kappa }\right) -\frac{1}{2}\xi ^{\alpha }\delta _{\alpha \beta \gamma \delta }^{\mu\sigma \eta \tau }B_{~\kappa \eta \tau }^{\gamma }D_{\mu }\left( \hat{g}^{\beta \rho }g^{\delta \kappa }\right)  \\
& =  ~\xi ^{\alpha }\delta _{\alpha \beta \gamma \delta }^{\mu \left[ \rho\right\vert \eta \tau }\hat{h}_{\kappa \lambda }^{\beta \left\vert \sigma \right]\delta \epsilon }B_{~\epsilon \eta \tau }^{\gamma }Q_{\mu }^{~\kappa \lambda},
\end{split}
\label{100}
\end{equation}
where%
\begin{equation}
\hat{h}_{\kappa \lambda }^{\beta \sigma \delta \epsilon }\equiv -\frac{1}{2}\hat{g}%
^{\beta \sigma }g^{\delta \epsilon }g_{\kappa \lambda }+\hat{g}^{\delta
\epsilon }\delta _{\left( \kappa \right. }^{\beta }\delta _{\left. \lambda
\right) }^{\sigma }+\hat{g}^{\beta \sigma }\delta _{\left( \kappa \right.
}^{\delta }\delta _{\left. \lambda \right) }^{\epsilon }.  \label{101}
\end{equation}%
With these definitions, eq.~(\ref{98}) assumes the form%
\begin{equation}
\hat{J}_{\mathcal{P}}^{\rho }=\hat{W}_{\text{GL}}^{\rho }+\partial _{\sigma }\left( \hat{U}%
_{\mathcal{P}}^{\prime \rho \sigma }+\hat{Z}^{\rho \sigma }\right) .
\label{102}
\end{equation}%
The terms inside the divergence are antisymmetric in $\rho $ and $\sigma $, so that $\hat{X}_{\mathcal{P}}=\partial_{\rho}\hat{J}_{\mathcal{P}}^{\rho }=\partial_{\rho}\hat{W}_{\text{GL}}^{\rho }$.

Note that the representation in eq.~(\ref{102}) preserves the symmetry of the charge, in the sense that $\hat{X}_{\mathcal{P}}=0$ on-shell. To see this, we observe that with the vanishing of non-metricity and its higher-order derivatives on the boundary as well as the variational derivatives, the divergence of the current reduces simply to $\partial_{\rho}\hat{W}_{\text{GL}}^{\rho }=-2\xi^{\mu}g^{\rho \nu}D_{\rho}\hat{\mathcal{L}}_{\mu \nu}=0$ on the boundary. The latter is a property of the Einstein tensor and its generalizations \cite{Lo}.

A variation of the current yields%
\begin{equation}
\delta \hat{J}_{\mathcal{P}}^{\rho }=\frac{\delta \hat{W}_{\text{GL}}^{\rho }}{\delta y_{A}%
}\delta y_{A}+\partial _{\sigma }\left( \frac{\partial \hat{W}_{\text{GL}%
}^{\rho }}{\partial \partial _{\sigma }y_{A}}\delta y_{A}+\delta \hat{U}_{%
\mathcal{P}}^{\prime \rho \sigma }+\delta \hat{Z}^{\rho \sigma }\right) .
\label{103}
\end{equation}%
Since $\hat{Z}^{\rho \sigma}$ contains only first-order derivatives, its variation may be written as
\begin{equation}
\delta \hat{Z}^{\rho \sigma}=\frac{\partial \hat{Z}^{\rho \sigma }}{%
\partial y_{A}}\delta y_{A}+\frac{\partial \hat{Z}^{\rho \sigma }}{\partial
\partial _{\lambda }y_{A}}\partial _{\lambda }\delta y_{A}.
\label{104}
\end{equation}%
Then the variation of the current becomes:%
\begin{equation}
\delta \hat{J}^{\rho}_{\mathcal{P}}=\frac{\delta \hat{W}_{\text{GL}}^{\rho }}{\delta y_{A}}\delta y_{A} + \partial _{\sigma }\left( \frac{\partial \hat{W}_{\text{GL}}^{\rho }}{\partial \partial _{\sigma }y_{A}}\delta y_{A}+\frac{\partial \hat{Z}^{\rho\sigma }}{\partial y_{A}}\delta y_{A}+\frac{\partial \hat{Z}^{\rho \sigma }}{\partial \partial _{\lambda }y_{A}}\partial _{\lambda }\delta y_{A}+\delta\hat{U}_{\mathcal{P}}^{\prime \rho \sigma }\right) .
\label{105}
\end{equation}%
Now as before, the guiding principle is to eliminate the divergence terms in eq.~(\ref{105}). Similarly to eq.~(\ref{62}), the symmetric part of the first-term in the divergence does not contribute, so we may drop it completely. The new equation for the variation of the superpotential becomes
\begin{equation}
\left. \delta \hat{U}_{\mathcal{P}}^{\prime \rho \sigma }\right\vert
_{\partial \mathcal{M}}=-\frac{\partial \hat{W}_{\text{GL}}^{\left[ \rho
\right. }}{\partial \partial _{\left. \sigma \right] }y_{A}}\delta y_{A}-%
\frac{\partial \hat{Z}^{\rho \sigma }}{\partial y_{A}}\delta y_{A}-\frac{%
\partial \hat{Z}^{\rho \sigma }}{\partial \partial _{\lambda }y_{A}}\partial
_{\lambda }\delta y_{A}.  \label{106}
\end{equation}%
Eq.~(\ref{106}) must be supplemented by boundary conditions. Assuming the boundary conditions in (\ref{86}), and substituting eq.~(\ref{88}) into the third term in the r.h.s. of eq.~(\ref{106}), we find:%
\begin{equation}
\left. \delta \hat{U}_{\mathcal{P}}^{\prime \rho \sigma }\right\vert_{\partial \mathcal{M}} = -\frac{\partial \hat{W}_{\text{GL}}^{\left[ \rho\right. }}{\partial \partial _{\left. \sigma \right] }\Gamma _{~\mu \nu}^{\lambda }}\delta \Gamma _{~\mu \nu }^{\lambda }-\frac{\partial \hat{Z}^{\rho \sigma }}{\partial \Gamma _{~\mu \nu }^{\lambda }}\delta \Gamma_{~\mu \nu }^{\lambda } + 2g^{\kappa \left( \mu \right. }\delta _{\eta }^{\left. \nu \right) }\frac{\partial \hat{Z}^{\rho \sigma }}{\partial \partial _{\lambda }g^{\mu \nu }} \delta \Gamma _{~\kappa \lambda }^{\eta }.
 \label{107}
\end{equation}%
The first term yields $\hat{U}^{\rho \sigma}$ in eq.~(\ref{64}), while the second and third terms in eq.~(\ref{107}) cancel each other out (see Appendix for details). We conclude that $\hat{Z}^{\rho \sigma}$ does not contribute to the superpotential.

Motivated by the generalized Bianchi identity in eq.~(\ref{21}), we have shown that it is possible to split the Palatini current into first and second-order contributions. Applying the variational principle to the current, we have found this splitting produces a superpotential, $\hat{U}_{\mathcal{P}}^{\prime \rho \sigma }$, that would be obtained from $\hat{U}^{\rho \sigma }$ upon symmetry breaking and gauge fixing at the boundary. This is at odds with $\hat{U}_{\mathcal{P}}^{\rho \sigma }$ in eq.~(\ref{91}), but unsurprisingly perhaps, is commensurate with the general linear formulation.

We are led to conclude that even within the Palatini formulation, it is possible to obtain two different superpotentials by a reshuffling of the current density! Furthermore, these results imply that the KL equation, given in eq.~(\ref{85}), is more general than the JS equation, eq.~(\ref{64}) (so named in respect to its originators) and consequently the KL superpotential, eq.~(\ref{91}), is more general than the JS superpotential, eq.~(\ref{72}). It seems that both the JS current and the JS superpotential are contained within the Palatini formulation, but that the Palatini formulation is more general.

We must make here a \textit{caveat}. The splitting of the covariant derivative of the variational derivative with respect to the affine connection into second-order derivatives and first-order terms is not necessarily valid for an arbitrary Lagrangian density with first-order fields equations. Furthermore, the splitting in eq.~(\ref{93}) is not unique. It is possible to continue this splitting further, to extract first-order terms by further antisymmetrization of the covariant derivatives in $D_{\sigma }\hat{\mathcal{L}}_{\lambda }^{~\rho \sigma }$.

\section{\label{sec:level4}The Renormalization of the Palatini Lagrangian}

Comparing both superpotentials, eqs.~(\ref{72}) and (\ref{91}), we have found that they differ when the curvature is coupled to another piece of curvature. It appears that the bulk non-metricity plays a pivotal role in the Palatini formulation, and generates a non-zero contribution to the superpotential, even though it, its variation and its higher-order derivatives vanish on the boundary. Hence, even though non-metricity is present in the bulk of both $GL(N,\mathbf{R})$ and Palatini space-times, only the latter is affected by it. This behaviour is reminiscent of the role that ghosts play in non-Abelian QFT. The non-metricity tensor, introduced initially as an auxiliary field to maintain first-order variational derivatives, should have left no trace on the physical charge.

In this section, we explore this conundrum in greater detail. We discover that the difference between the superpotentials in the Palatini and the general linear formulations, $\hat{U}^{\rho \sigma}_{\mathcal{P}}$ and $\hat{U}^{\rho \sigma}$ respectively, stems from a Lagrangian density that explicitly couples quadratic terms in the non-metricity with the curvature tensor. We show that this Lagrangian density may be generalized to a whole class of Lagrangian densities that may be added, \emph{ex post facto}, to the standard Lovelock Lagrangians to remove the additional terms that arise from non-metricity in the Palatini formulation. Demonstrably, its inclusion in the $GL(N,\mathbf{R})$ formulation does not alter $\hat{U}^{\rho \sigma}$.

This removal, however, is quite artificial. For instance, it is not clear, \emph{a priori}, what the coupling constant (value and sign) should be. These values are deduced only after the explicit calculation of $\hat{U}^{\rho \sigma}_{\mathcal{P}}$ is completed. Furthermore, we may examine the problem from a very different point of view. If the non-metricity in the bulk can contribute to the conserved charges on the boundary, then perhaps we should endeavour to treat it not as an auxiliary field, but as a real physical field that carries with it some mass, spin or radiation. This is not a new idea \cite{P,Baekler}, but we believe its implementation in this context is quite new.

In this spirit we present a parity-violating Lagrangian density that extends the superpotential into asymptotically maximally symmetric space-times. We consider its contribution in two cases: a $4$-dimensional solution of an asymptotically Kerr-AdS space-time proposed by Henneaux and Teitelboim \cite{HT}, and a $4$-dimensional mass-less gravo-magnetic analogue, motivated by the work of Lynden-Bell and Nouri-Zonoz \cite{LBNZ}.

\subsection{Explicit Non-metricity}

We begin by introducing the following Lagrangian density:
\begin{equation}
\hat{\mathcal{L}}_{\mathrm{i}}^{\mathcal{Q}}\equiv \hat{g}_{\rho \sigma}\delta _{\kappa \lambda }^{\mu \nu }Q_{\mu }^{~\kappa \rho }Q_{\nu
}^{~\lambda \sigma },  \label{108}
\end{equation}%
where $\mathcal{Q}$ denotes Lagrangians with explicit non-metricity, and may be used as a superscript or subscript or in parenthesis for the sake of clarity.

$\hat{\mathcal{L}}_{\mathrm{i}}^{\mathcal{Q}}$ has a unique provenance. We have found that it can be obtained from the work by Floreanini and Percacci \cite{FP}. Working within the $GL(4,\mathbf{R})$ formulation, the authors proved that both the metric-compatible and the torsion-free constraints naturally arose from the dynamics of $\hat{\mathcal{L}}_{1}$ (see eq.~(\ref{4})). To this end, they split $\hat{\mathcal{L}}_{1}$ into two parts: the Einstein-Hilbert part (with the Levi-Civita connection and second-order field equations) and a second part, that was quadratic in torsion and non-metricity. The latter, embodied in their eqs. $(2.6)$ and $(2.7b)$, can be used to obtain our eq.~(\ref{108}) by gauge fixing, eq.~(\ref{45}).

In addition to the obvious diffeomorphism invariance, $\hat{\mathcal{L}}_{\mathrm{i}}^{\mathcal{Q}}$ is also invariant under the volume-preserving transformation of the
affine connection \cite{Sm}, which reads%
\begin{equation}
\delta _{q}\Gamma _{~\mu \nu }^{\lambda }\equiv -\frac{1}{2}\left( \delta_{\mu }^{\lambda }Q_{\nu }+\delta _{\nu }^{\lambda }Q_{\mu }-g_{\mu \nu}g^{\kappa \lambda }Q_{\kappa }\right) .
\label{109}
\end{equation}%
Here $Q_{\nu }\equiv g_{\alpha \beta }D_{\nu }g^{\alpha \beta }$ is the Weyl vector and consequently $\delta _{q}Q_{\lambda }^{~\mu \nu }=-g^{\mu \nu}Q_{\lambda }$. In fact, this invariance is more general, and $Q_{\nu }$ can be any covariant vector field. Note that the non-metricity that appears in eq.~(\ref{108}) can be replaced by the trace-free non-metricity, defined as $\slashed{Q}_{\lambda }^{~\mu \nu }\equiv Q_{\lambda }^{~\mu \nu }-\frac{1}{N}
g^{\mu \nu}Q_{\lambda}$. Eq. (\ref{108}) lends itself naturally to the following generalization to a new class of Lagrangian densities with quadratic non-metricity and first-order field equations:
\begin{equation}
\hat{\mathcal{L}}_{\mathrm{ii}}^{\mathcal{Q}}=\hat{g}_{\rho \sigma }\delta
_{\kappa \lambda \alpha \beta }^{\mu \nu \gamma \delta }B_{~~\gamma \delta
}^{\alpha \beta }Q_{\mu }^{~\kappa \rho }Q_{\nu }^{~\lambda \sigma }.
\label{110}
\end{equation}%
Higher orders are considered in the Appendix. Note that $\hat{\mathcal{L}}_{\mathrm{ii}}^{\mathcal{Q}}$ is no longer volume-invariant. The variational
derivatives are given by:
\begin{equation}
\frac{1}{\sqrt{|g|}}\frac{\delta \hat{\mathcal{L}}_{\mathrm{i}}^{\mathcal{Q}}%
}{\delta \Gamma _{~\rho \sigma }^{\lambda }}=2h_{\mu \nu \lambda }^{\rho
\sigma \kappa }Q_{\kappa }^{~\mu \nu },  \label{111}
\end{equation}%
where%
\begin{equation}
h_{\mu \nu \lambda }^{\rho \sigma \kappa }\equiv g^{\rho \sigma }\delta_{\left( \mu \right. }^{\kappa }g_{\left. \nu \right) \lambda }-\delta
_{\lambda }^{\kappa }\delta _{\left( \mu \right. }^{\rho }\delta _{\left.\nu \right) }^{\sigma } +\delta _{~\lambda }^{\left( \rho \right. }\delta _{\left( \mu \right.}^{\left. \sigma \right) }\delta _{\left. \nu \right) }^{\kappa }-g^{\kappa\left( \rho \right. }\delta _{\left( \mu \right. }^{\left. \sigma \right)
}g_{\left. \nu \right) \lambda };
\label{112}
\end{equation}%
and%
\begin{equation}
\frac{1}{\sqrt{|g|}} \frac{\delta \hat{\mathcal{L}}_{\mathrm{i}}^{\mathcal{Q}}}{\delta g^{\rho \sigma }}=-2g_{\kappa \left( \rho \right. }\delta
_{\sigma )\lambda }^{\mu \nu }D_{\mu }Q_{\nu }^{~\kappa \lambda } + Q_{\mu }^{~\kappa \eta }Q_{\nu }^{~\lambda \tau }\left( -\delta _{\kappa
\lambda }^{\mu \nu }h_{\rho \sigma \eta \tau }+2\delta _{\lambda \left( \rho \right. }^{\mu \nu }h_{\left. \sigma \right) \tau \kappa \eta }\right) ,
\label{113}
\end{equation}%
where%
\begin{equation}
h_{\mu \nu \rho \sigma }\equiv \frac{1}{2}g_{\mu \nu }g_{\rho \sigma}+g_{\mu \left( \rho \right. }g_{\left. \sigma \right) \nu }.
\label{114}
\end{equation}%
We see that the variational derivative with respect to the metric, $\delta_{\mathbf{g}}\hat{\mathcal{L}}^{\mathcal{Q}}_{\mathrm{i}}\sim \mathbf{g\cdot B}+\mathbf{Dg\cdot Dg}$. Our boundary conditions, eq.~(\ref{86}), dictate that on the boundary the non-metricity tensor should vanish. This means that the quadratic terms $\mathbf{Dg\cdot Dg}$ cannot contribute to the superpotential on the boundary, and the only relevant part is the first term, which can be rewritten using eq.~(\ref{31}) in terms of the curvature tensor:
\begin{equation}
\frac{1}{\sqrt{|g|}}\frac{\delta \hat{\mathcal{L}}_{\mathrm{i}}^{\mathcal{Q}}%
}{\delta g^{\rho \sigma }}\overset{\bullet }{=}-4g_{\mu \left( \rho \right.
}B_{~~~\left. \sigma \right) \nu }^{\left( \mu \nu \right) }.  \label{115}
\end{equation}%
Substituting eqs.~(\ref{111}) and (\ref{115}) into eq.~(\ref{90}), we obtain
expressions for $\hat{Y}^{\rho \sigma }$ and $\hat{W}_{\mathcal{P}}^{\rho }$,%
\begin{equation}
\hat{W}_{\mathcal{P}}^{\rho }\overset{\bullet }{=}8\xi ^{\left( \mu \right.
}g^{\left. \nu \right) \rho }g_{\kappa \mu }B_{~~~~\nu \lambda }^{\left(
\kappa \lambda \right) }~\text{and}~\hat{Y}^{\rho \sigma }=2\xi ^{\lambda
}h_{\mu \nu \lambda }^{\rho \sigma \kappa }Q_{\kappa }^{~\mu \nu },
\label{116}
\end{equation}%
which we differentiate according to eq.~(\ref{89}). We find%
\begin{equation}
-\frac{4}{3}\frac{\partial \hat{Y}^{\lambda \left[ \rho \right. }}{\partial \left( \partial _{\left. \sigma \right] }g^{\mu \nu }\right) }
g^{\kappa \left( \mu \right. }\Delta _{~~\kappa \lambda }^{\left. \nu \right) }=
 -2\left( \xi ^{\left[ \rho \right. }g^{\left. \sigma \right] \mu }\Delta_{~\mu \nu }^{\nu }+\xi ^{\mu }g^{\nu \left[ \rho \right. }\Delta _{~\mu \nu
}^{\left. \sigma \right] }+g^{\mu \nu }\xi ^{\left[ \rho \right. }\Delta_{~\mu \nu }^{\left. \sigma \right] }\right),
\label{117}
\end{equation}%
and%
\begin{equation}
-\frac{\partial \hat{W}_{\mathcal{P}}^{\left[ \rho \right. }}{\partial \left( \partial_{\left. \sigma \right] }\Gamma _{~\mu \nu }^{\lambda }\right) }\Delta
_{~\mu \nu }^{\lambda }= 2\left( g^{\mu \nu }\xi ^{\left[ \rho \right. }\Delta _{\text{ \ }\mu \nu}^{\left. \sigma \right] }+\xi ^{\mu }g^{\nu \left[ \rho \right. }\Delta_{~~\mu \nu }^{\left. \sigma \right] }+\xi ^{\left[ \rho \right. }g^{\left.\sigma \right] \nu }\Delta _{~\mu \nu }^{\mu }\right).
\label{118}
\end{equation}
Substituting eqs.~(\ref{117}) and (\ref{118}) into eq.~(\ref{89}), we find that
\begin{equation}
\hat{U}^{\rho \sigma}_{\mathrm{i}}=-\frac{\partial \hat{W}_{\mathcal{P}}^{\left[ \rho \right. }}{\partial \left( \partial
_{\left. \sigma \right] }\Gamma _{~\mu \nu }^{\lambda }\right) }\Delta
_{~\mu \nu }^{\lambda }-\frac{4}{3}\frac{\partial \hat{Y}^{\lambda \left[
\rho \right. }}{\partial \left( \partial _{\left. \sigma \right] }g^{\mu \nu
}\right) }g^{\kappa \left( \mu \right. }\Delta _{~~\kappa \lambda }^{\left.
\nu \right) }=0.  \label{119}
\end{equation}%
Hence $\hat{\mathcal{L}}_{\mathrm{i}}^{\mathcal{Q}}$ does not contribute to the superpotential on the boundary! It seems, at least from eq.~(\ref{119}), that the parity-preserving form of quadratic non-metricity by itself cannot contribute to the superpotential (compare, however, with eqs.~(\ref{148a})-(\ref{148b})).

Nevertheless, $\hat{\mathcal{L}}_{\mathrm{ii}}^{\mathcal{Q}}$ already couples curvature and non-metricity. Its variational derivative with respect to
metric reads:
\begin{equation}
\begin{split}
\frac{1}{\sqrt{|g|}} \frac{\delta \hat{\mathcal{L}}_{\mathrm{ii}}^{\mathcal{%
Q}}}{\delta g^{\rho \sigma }} = & -2g_{\kappa \left( \rho \right. }\delta
_{\sigma )\lambda \alpha \beta }^{\mu \nu \gamma \delta }B_{~~\gamma \delta
}^{\alpha \beta }D_{\mu }Q_{\nu }^{~\kappa \lambda } \\
& +B_{~\varepsilon \gamma \delta }^{\alpha }Q_{\eta }^{~\kappa \eta }Q_{\nu
}^{~\lambda \tau }\left( \delta _{\kappa \lambda \alpha \beta }^{\mu \nu
\gamma \delta }f_{\rho \sigma \eta \tau }^{\beta \varepsilon }+2\delta
_{\alpha \beta \lambda \left( \rho \right. }^{\mu \nu \gamma \delta
}f_{\sigma )\tau \kappa \eta }^{\beta \varepsilon }\right) ,
\end{split}
\label{120}
\end{equation}%
where%
\begin{equation}
f_{\rho \sigma \mu \nu }^{\kappa \lambda }\equiv -g^{\kappa \lambda }h_{\mu
\nu \rho \sigma }+g_{\mu \nu }\delta _{\left( \rho \right. }^{\kappa }\delta
_{\left. \sigma \right) }^{\lambda }.
\label{121}
\end{equation}%
We see that $\delta_{\mathbf{g}}\hat{\mathcal{L}}^{\mathcal{Q}}_{\mathrm{ii}}\sim \mathbf{g\cdot B\cdot B}+\mathbf{g\cdot Dg\cdot Dg\cdot B}$. The latter terms do not contribute to the superpotential on the boundary. So the relevant part becomes%
\begin{equation}
\frac{\delta \hat{\mathcal{L}}_{\mathrm{ii}}^{\mathcal{Q}}}{\delta g^{\rho
\sigma }}\overset{\bullet }{=}-2\hat{g}_{\kappa \left( \rho \right. }\delta
_{\left. \sigma \right) \lambda \alpha \beta }^{\mu \nu \gamma \delta
}B_{~~\gamma \delta }^{\alpha \beta }B_{~~~~\mu \nu }^{\left( \kappa \lambda
\right) }.
\label{122}
\end{equation}
Substituting eq.~(\ref{122}) into $\hat{W}_{\mathcal{P}}^{\rho }$ in eq.~(%
\ref{90}), we have
\begin{equation}
\hat{W}_{\mathcal{P}}^{\rho }\overset{\bullet }{=}4\xi ^{\left( \mu \right.
}g^{\left. \nu \right) \rho }\hat{g}_{\kappa \mu }\delta _{\nu \lambda
\alpha \beta }^{\eta \tau \gamma \delta }B_{~~\gamma \delta }^{\alpha \beta
}B_{~~~\eta \tau }^{\left( \kappa \lambda \right) },
\label{123}
\end{equation}%
and, evaluated at the boundary, we find%
\begin{equation}
-\frac{\partial \hat{W}_{\mathcal{P}}^{\left[ \rho \right. }}{\partial \left(\partial _{\left. \sigma \right] }\Gamma _{~\mu \nu }^{\lambda }\right) }
\Delta _{~\mu \nu }^{\lambda } =-8\hat{g}_{\kappa \mu }\xi ^{\left( \mu \right. }g^{\left. \nu \right)\left[ \rho \right. }\delta _{\nu \lambda \alpha \beta }^{\sigma ]\tau \gamma \delta }R_{~~\gamma \delta }^{\alpha \beta }g^{\eta \left( \kappa \right. }\Delta _{~~\tau \eta }^{\left. \lambda \right) }.
\label{124}
\end{equation}%
The variational derivative with respect to the affine connection is%
\begin{equation}
\begin{split}
\frac{1}{\sqrt{|g|}}\frac{\delta \hat{\mathcal{L}}_{\mathrm{ii}}^{\mathcal{Q}%
}}{\delta \Gamma _{~\rho \sigma }^{\lambda }}=& ~2\delta _{\kappa \lambda
\mu \nu }^{\alpha \beta \gamma \left( \rho \right. }f_{\eta \tau \theta
\varphi }^{\left. \sigma \right) \kappa }Q_{\alpha }^{~\mu \eta }Q_{\beta
}^{~\nu \tau }Q_{\gamma }^{~\theta \varphi }+4\delta _{\lambda }^{\left( \eta \right. }g^{\left. \kappa \right) \left(
\rho \right. }\delta _{\kappa \nu \alpha \beta }^{\left. \sigma \right)
\varepsilon \gamma \delta }g_{\mu \eta }B_{~~\gamma \delta }^{\alpha \beta
}Q_{\varepsilon }^{~\mu \nu } \\
& +4g_{\varepsilon \chi }g^{\beta \left( \rho \right. }\delta _{~\kappa \tau
\lambda \beta }^{\left. \sigma \right) \eta \gamma \delta }B_{~~~\gamma
\delta }^{\left( \kappa \varepsilon \right) }Q_{\eta }^{~\tau \chi }.
\end{split}
\label{125}
\end{equation}%
The only relevant term that may contribute at boundary is the second one%
\begin{equation}
\frac{1}{\sqrt{|g|}}\frac{\delta \hat{\mathcal{L}}_{\mathrm{ii}}^{\mathcal{Q}%
}}{\delta \Gamma _{~\rho \sigma }^{\lambda }}\overset{\bullet }{=}4g_{\mu
\eta }\delta _{\lambda }^{\left( \eta \right. }g^{\left. \tau \right) \left(
\rho \right. }\delta _{\tau \nu \alpha \beta }^{\sigma )\kappa \gamma \delta
}B_{~~\gamma \delta }^{\alpha \beta }Q_{\kappa }^{~\mu \nu },  \label{126}
\end{equation}%
so that
\begin{equation}
\hat{Y}^{\rho \sigma }\overset{\bullet }{=}4\hat{g}_{\mu \eta }\xi ^{\left(
\eta \right. }g^{\left. \kappa \right) \left( \rho \right. }\delta _{~\kappa
\nu \alpha \beta }^{\left. \sigma \right) \lambda \gamma \delta }B_{~~\gamma
\delta }^{\alpha \beta }Q_{\lambda}^{~\mu \nu}.  \label{127}
\end{equation}%
From eq.~(\ref{127}) we find
\begin{equation}
-\frac{4}{3}\frac{\partial \hat{Y}^{\lambda \left[ \rho \right. }}{\partial \left( \partial _{\left. \sigma \right] }g^{\mu \nu }\right)}g^{\kappa \left( \mu \right. }\Delta _{~~\kappa \lambda }^{\left. \nu \right) }= -\frac{8}{3}\left( \xi ^{\left( \mu \right. }g^{\left. \nu \right) \tau}\delta _{\nu \lambda \alpha \beta }^{\rho \sigma \gamma \delta }-\xi^{\left( \mu \right. }g^{\left. \nu \right) \left[ \rho \right. }\delta
_{\nu \lambda \alpha \beta }^{\left. \sigma \right] \tau \gamma \delta
}\right) \hat{g}_{\mu \kappa }g^{\eta \left( \kappa \right. }\Delta _{~~\tau
\eta }^{\left. \lambda \right) }R_{~~\gamma \delta }^{\alpha \beta }.
\label{128}
\end{equation}
Summing up eqs.~(\ref{124}) and (\ref{128}), we obtain:%
\begin{equation}
\hat{U}_{\mathrm{ii}}^{\rho \sigma} = -\frac{8}{3}\hat{g}_{\kappa \mu }g^{\eta \left( \kappa \right. }\Delta_{~~\tau \eta }^{\left. \lambda \right) }R_{~~\gamma \delta }^{\alpha \beta} \left[ \xi ^{\left( \mu \right. }g^{\left. \nu \right) \tau }\delta _{\nu \lambda \alpha \beta }^{\rho \sigma \gamma \delta }+2\xi ^{\left( \mu \right. }g^{\left. \nu \right) \left[ \rho \right. }\delta _{\nu \lambda \alpha \beta }^{\left. \sigma \right] \tau \gamma \delta }\right].
\label{129}
\end{equation}%
Now we make use of the following identity%
\begin{equation}
\left( 2\xi ^{\left( \eta \right. }g^{\left. \tau \right) \left[ \rho \right. }\delta _{~\tau \lambda \alpha \beta }^{\left. \sigma \right] \mu
\gamma \delta }+\xi ^{\left( \eta \right. }g^{\left. \tau \right) \mu }\delta _{\tau \lambda \alpha \beta }^{\rho \sigma \gamma \delta }\right)
R_{~~\gamma \delta }^{\alpha \beta }\hat{g}_{\kappa \eta }g^{\nu \left(\kappa \right. }\Delta _{~~\mu \nu }^{\left. \lambda \right) }
 =\xi ^{\tau }\delta _{\tau \lambda \alpha \beta }^{\rho \sigma \gamma \mu}R_{~~\gamma \kappa }^{\alpha \beta }\hat{g}^{\nu \left( \kappa \right.
}\Delta _{~~\mu \nu }^{\left. \lambda \right) }.
\label{130}
\end{equation}%
Replacing eq.~(\ref{130}) into eq.~(\ref{129}), we have
\begin{equation}
\hat{U}_{\mathrm{ii}}^{\rho \sigma}=\frac{8}{3}\bar{\xi} ^{\alpha }\delta _{\alpha \beta \gamma \delta }^{\mu \nu
\rho \sigma }R_{~~\lambda \mu }^{\gamma \delta }\bar{\hat{g}}^{\kappa \left(\lambda \right. }\Delta _{~~\nu \kappa }^{\left. \beta \right) }.
\label{131}
\end{equation}
Comparing eq.~(\ref{131}) with eq.~(\ref{91}), we find that $\hat{U}_{\mathrm{ii}}^{\rho \sigma }=4{\Delta\hat{U}}_{2}^{\rho \sigma}$! Subsequently, we can re-normalize $\hat{\mathcal{L}}_{2}^{\mathcal{P}}$ in eq.~(\ref{23}) as follows
\begin{equation}
\hat{\mathcal{L}}_{2}^{\prime \mathcal{P}}\equiv \frac{1}{4}\delta _{\alpha\beta \gamma \delta }^{\mu \nu \rho \sigma }\left( \hat{B}_{~~\mu \nu}^{\alpha \beta }-\hat{g}_{\kappa \lambda }Q_{\mu }^{~\alpha \kappa }Q_{\nu}^{~\beta \lambda }\right) B_{~~\rho \sigma }^{\gamma \delta },  \label{132}
\end{equation}
where a prime, as in $\hat{\mathcal{L}}^{\prime }$, denotes a re-normalized Lagrangian density. The superpotential obtained from $\hat{\mathcal{L}}_{2}^{\prime \mathcal{P}}$ by means of eq.~(\ref{89}) is now simply ${\hat{U}}_{2}^{\rho \sigma }$ in eq.~(\ref{72}), and the discrepancy between the formulations is resolved by taking
\begin{equation}
\hat{\mathcal{L}}_{2}^{\prime }=\frac{1}{4}\sqrt{\left\vert \gamma \right\vert }\theta _{abcd}^{\mu \nu \rho \sigma }\left( B_{~~\mu \nu
}^{ab}-\gamma _{kl}\left( \nabla _{\mu }\gamma ^{ak}\right) \left( \nabla_{\nu }\gamma ^{bl}\right) \right) B_{~~\rho \sigma }^{cd}.  \label{133}
\end{equation}

\subsection{Boundary terms and the coupling of $\hat{\mathcal{L}}_{\mathrm{i}}^{\mathcal{Q}}$}

The renormalized Lagrangian density, $\hat{\mathcal{L}}_{2}^{\prime \mathcal{P}}$ in eq.~(\ref{132}), has fixed the numerical coefficient or coupling strength of $\hat{\mathcal{L}}_{\mathrm{ii}}^{\mathcal{Q}}$. In order to remove ${\Delta\hat{U}}_{2}^{\rho \sigma }$, we have found this value to be $-1/4$. However, these considerations cannot determine the coupling of $\hat{\mathcal{L}}_{\mathrm{i}}^{\mathcal{Q}}$, since $\hat{U}_{\mathrm{i}}^{\rho \sigma }=0$. Yet, this arbitrariness is misleading. To solve this riddle, we look elsewhere, by examining boundary and surface terms.

A compact space-time manifold, $\mathcal{M}$, with a boundary, $\partial \mathcal{M}$, usually mandates changes to the action, which take into account boundary terms. Typically, in order to formulate a well-defined variational principle, Dirichlet boundary conditions on the metric require the introduction of surface terms to the action \cite{York,Obukhov} to remove variations of the affine connection. Such terms become necessary when a divergence appears in the variation, which cannot be reconciled with the field equations. This is true in particular for Lovelock Lagrangians, but is not strictly necessary in the case of $\hat{\mathcal{L}}_{\mathrm{n}}^{\mathcal{Q}}$, as the boundary terms vanish by virtue of the vanishing of non-metricity on the boundary, which satisfies the field equations as well.

Nonetheless, here we examine the boundary term that arises upon the variation of $\hat{\mathcal{L}}_{\mathrm{i}}^{\mathcal{Q}}$. We find that it can be used to complete the surface term that was proposed by Obukhov \cite{Obukhov}, whose goal was to reformulate the Palatini Lagrangian with a boundary in four dimensions.

Obukhov \cite{Obukhov} defines the following scalar
\begin{equation}
K \equiv \frac{1}{2} \left( D_{\mu} n^{\mu} +D^{\mu} n_{\mu} \right)=\frac{1}{2} \left( Q^{~\mu \nu}_{\mu}n_{\nu}+2g^{\mu \nu}D_{\mu}n_{\nu} \right),
\label{199}
\end{equation}
where $n^{\mu}$ is the unit vector normal to the boundary $\partial \mathcal{M}$. For simplicity we work in $N=4$ dimensions. The normalization is given by $n^{\mu} n_{\mu}=\varepsilon$, where $\varepsilon=\pm 1$ for a space-like or time-like hypersurface, respectively. The variation of $n_{\nu}$ is given by $\delta n_{\nu}=-\frac{1}{2}\varepsilon n_{\nu} n_{\alpha} n_{\beta} \delta g^{\alpha \beta}$. We define the projection tensor, $h^{\mu}_{~\nu}\equiv \delta^{\mu}_{\nu}-\varepsilon n^{\mu}n_{\nu}$, and with it the induced 3-metric $h_{\mu \nu}=g_{\mu \nu}-\varepsilon n_{\mu} n_{\nu}$. Indices are raised and lowered with $\mathbf{g}$.

$K$ is closely related to the extrinsic curvature, and in the metric case, $Q_{\lambda}^{~\mu \nu}=0$, reduces to the trace of the second fundamental form. It is chosen so that the variations of the affine connection on the boundary would cancel their corresponding part in $\delta \hat{\mathcal{L}}_{1}^{\mathcal{P}}$ (see eq.~(\ref{23})), given by
\begin{equation}
\delta \hat{\mathcal{L}}_{1}^{\mathcal{P}}=\frac{\delta \hat{\mathcal{L}}_{1}^{\mathcal{P}}}{\delta g^{\mu \nu}}\delta g^{\mu \nu}+\frac{\delta \hat{\mathcal{L}}_{1}^{\mathcal{P}}}{\delta \Gamma^{\lambda}_{~\mu \nu}}\delta \Gamma^{\lambda}_{~\mu \nu}+\partial_{\mu}{\hat{k}^{\mu}}.
\label{200}
\end{equation}
The variational derivatives are given in eq.~(\ref{27}), and
\begin{equation}
\hat{k}^{\mu}\equiv \hat{g}^{\kappa \lambda} \delta \Gamma^{\mu}_{~\kappa \lambda} - \hat{g}^{\lambda \mu} \delta \Gamma^{\kappa}_{~\kappa \lambda}.
\label{201}
\end{equation}
Under the sign of the integral, we have
\begin{equation}
\sqrt{\left\vert h \right\vert} n_{\mu} k^{\mu}=-\sqrt{\left\vert h \right\vert }g^{\mu \nu} \left( n_{\nu}\delta \Gamma^{\lambda}_{~\lambda \mu} - n_{\lambda} \delta \Gamma^{\lambda}_{~\mu \nu} \right).
\label{202}
\end{equation}
The variation of $2K$ is given by
\begin{equation}
\begin{split}
2\delta K = & ~g^{\mu \nu} \left( n_{\nu} \delta \Gamma^{\lambda}_{~\lambda \mu} - n_{\lambda} \delta \Gamma^{\lambda}_{~\mu \nu} \right) \\
&+\delta h^{\mu \nu} D_{\mu} n_{\nu} + h^{\mu}_{~\nu}D_{\mu} \left( h^{\nu}_{~\lambda} \delta n^{\lambda} \right) \\
&+\frac{\varepsilon}{2}n_{\beta}n_{\nu} \delta g^{\mu \nu} \left( n_{\mu}Q^{~\alpha \beta}_{\alpha}-n_{\alpha}Q_{\mu}^{~\alpha \beta}\right).
\end{split}
\label{203}
\end{equation}
A useful identity in the derivation of eq.~(\ref{203}) is $n^{\nu}D_{\mu}n_{\nu}=-\frac{1}{2}n_{\kappa}n_{\lambda}Q_{\mu}^{~\kappa \lambda}$. One can immediately see why this particular boundary term was chosen. Under the sign of the integral, eq.~(\ref{203}) multiplied by $\sqrt{\left\vert h \right\vert }$ and added to eq.~(\ref{202}) cancels the variations of $\Gamma^{\lambda}_{~\mu \nu}$, leaving only variations of the metric, which vanish on the boundary. More relevant to our cause is the third line in eq.~(\ref{203}). The explicit factor of $\varepsilon$ does not appear to reconcile with the fact that $K$ is a scalar, and should not depend on the nature of the boundary.

Similarly to eq.~(\ref{200}), for $\delta \hat{\mathcal{L}}_{\mathrm{i}}^{\mathcal{Q}}$ we have
\begin{equation}
\delta \hat{\mathcal{L}}_{\mathrm{i}}^{\mathcal{Q}}=\frac{\delta \hat{\mathcal{L}}_{\mathrm{i}}^{\mathcal{Q}}}{\delta g^{\mu \nu}}\delta g^{\mu \nu}+\frac{\delta \hat{\mathcal{L}}_{\mathrm{i}}^{\mathcal{Q}}}{\delta \Gamma^{\lambda}_{~\mu \nu}}\delta \Gamma^{\lambda}_{~\mu \nu}+\partial_{\mu}{\hat{v}^{\mu}}.
\label{204}
\end{equation}
The variational derivatives are given in eqs.~(\ref{111})-(\ref{114}), and
\begin{equation}
\hat{v}^{\mu}\equiv 2 \hat{g}_{\kappa \nu} Q_{\lambda}^{~\kappa \lambda}\delta g^{\mu \nu} - 2 \hat{g}_{\kappa \nu} Q_{\lambda}^{~\mu \nu} \delta g^{\kappa \lambda}.
\label{205}
\end{equation}
Under the sign of the integral, we have
\begin{equation}
\sqrt{\left\vert h \right\vert } n_{\mu} v^{\mu}=2 \sqrt{\left\vert h \right\vert } g_{\beta \nu} \delta g^{\mu \nu} \left( n_{\mu}Q_{\alpha}^{~\alpha \beta}- n_{\alpha} Q_{\mu}^{~\alpha \beta} \right).
\label{206}
\end{equation}
Eq.~(\ref{206}) is volume-invariant in the sense of eq.~(\ref{109}). Comparing eq.~(\ref{206}) with the expression in the third line in eq.~(\ref{203}), we see the similarity. The expressions in parenthesis are identical! This suggests that they may complement each other, by choosing  the coefficient for $\hat{\mathcal{L}}_{\mathrm{i}}^{\mathcal{Q}}$ to be $-1/4$. With this choice, under the integral sign, we obtain
\begin{equation}
\sqrt{\left\vert h \right\vert } \left(-\frac{1}{4} n_{\mu} v^{\mu}+2 \delta K \right) \overset{\bullet }{=} -\frac{1}{2} \hat{h}_{\beta \nu}\delta g^{\mu \nu} \left( n_{\mu}Q_{\alpha}^{~\alpha \beta}- n_{\alpha} Q_{\mu}^{~\alpha \beta} \right).
\label{207}
\end{equation}
With eq.~(\ref{207}) we can switch from $\delta g^{\mu \nu}$ to variations of $h_{\mu \nu}$, which is the genuine metric on the boundary. Thus, we have found an additional constraint, through $K$ in eq.~(\ref{199}), that binds together two distinct contributions. Consequently, we can define:
\begin{equation}
\hat{\mathcal{L}}_{1}^{\prime \mathcal{P}} \equiv \frac{1}{2} \delta^{\mu \nu}_{\kappa \lambda}
\left( \hat{B}^{\kappa \lambda}_{~~\mu \nu} - \frac{1}{2} \hat{g}_{\rho \sigma} Q_{\mu}^{~\kappa \rho}Q_{\nu}^{~\lambda \sigma} \right).
\end{equation}
This result agrees with the coupling of the $n$th Lagrangian density, $-n/2^{n+1}$,~$n \ge 1$. See Appendix for details.

It must be noted that $K$ is not unique, in the sense that it is possible to construct other boundary terms that would reduce to the second fundamental form when $Q_{\lambda}^{~\mu \nu}=0$.

\subsection{Symmetries of the superpotential equation}

The choice of dynamic variables is a cardinal one, because it dictates the dynamics of the system. In the preceding section we have shown that identical dynamics are insufficient in determining uniquely the superpotential. Indeed, the general linear and Palatini formulations have the same equations of motion on the boundary, but their respective superpotentials differ. Here we explore another puzzling facet of this question. We show below the surprising fact, that in the calculation of the superpotential for $\hat{\mathcal{L}}_{\mathrm{ii}}^{Q}$, we could choose as independent fields the metric tensor, the affine connection and the curvature tensor, and we would still end up with the same superpotential. This is a highly non-trivial fact. It implies a certain degeneracy exists in the choice of fields, with redundant dynamics which still yield the same superpotential.

In order to see this, we examine a more general Palatini formulation in which the curvature tensor, $B_{~\nu \rho \sigma }^{\mu }$, is independent of the affine connection. We define the following function%
\begin{equation}
F_{~\sigma \mu \nu }^{\rho }\left[ \mathbf{\Gamma} \right] \equiv 2\partial _{\left[
\mu \right. }\Gamma _{~\left. \nu \right] \sigma }^{\rho }+2\Gamma _{~\left[
\mu \right\vert \lambda }^{\rho }\Gamma _{~\left\vert \nu \right] \sigma
}^{\lambda },  \label{134}
\end{equation}%
for convenience. On the boundary, in addition to (\ref{86}), we require that $\left. B_{~\sigma \mu \nu }^{\rho }\right\vert _{\partial \mathcal{M}}=F_{~\sigma \mu \nu }^{\rho }$. We have
\begin{equation}
\hat{\mathcal{L}}_{\mathrm{ii}}^{\mathcal{Q}}\left[ g^{\mu \nu },\Gamma
_{~\mu \nu }^{\lambda },B_{~\nu \rho \sigma }^{\mu }\right] =\hat{g}_{\rho
\sigma }g^{\beta \eta }\delta _{\kappa \lambda \alpha \beta }^{\mu \nu
\gamma \delta }B_{~\eta \gamma \delta }^{\alpha }Q_{\mu }^{~\kappa \rho
}Q_{\nu }^{~\lambda \sigma }.  \label{135}
\end{equation}%
Clearly, $\hat{W}_{\mathcal{P}}^{\rho}$ is altered by this change, but since $\delta_{\mathbf{B}} \hat{\mathcal{L}}_{\mathrm{ii}}^{\mathcal{Q}} \sim \mathbf{g}\cdot \left(\mathbf{D}\mathbf{g}\right) \cdot \left( \mathbf{D}\mathbf{g}\right)$, it cannot contribute to the superpotential on the boundary. The relevant derivatives that do contribute are
\begin{eqnarray}
\frac{1}{\sqrt{|g|}}\frac{\delta \hat{\mathcal{L}}_{\mathrm{ii}}^{\mathcal{Q}%
}}{\delta \Gamma _{~\rho \sigma }^{\lambda }} & \overset{\bullet }{=} & 4g_{\mu
\eta }\delta _{\lambda }^{\left( \eta \right. }g^{\left. \tau \right) \left(
\rho \right. }\delta _{\tau \nu \alpha \beta }^{\sigma )\kappa \gamma \delta
}B_{~~\gamma \delta }^{\alpha \beta }Q_{\kappa }^{~\mu \nu },  \label{136} \\
\frac{1}{\sqrt{|g|}}\frac{\delta \hat{\mathcal{L}}_{\mathrm{ii}}^{\mathcal{Q}%
}}{\delta g^{\rho \sigma }}& \overset{\bullet }{=} & -2g_{\kappa \left( \rho
\right. }\delta _{\left. \sigma \right) \lambda \alpha \beta }^{\mu \nu
\gamma \delta }B_{~~\gamma \delta }^{\alpha \beta }F_{~~~\mu \nu }^{\left(
\kappa \lambda \right) }.  \label{137}
\end{eqnarray}
This result is not metric-compatible! In fact, all equations of motion of $\hat{\mathcal{L}}_{n}^{\mathcal{Q}}$ are trivially satisfied by the metricity condition on the boundary, which implies a certain redundancy exists in their solutions. Consequently, it implies that the explicit curvature in $\hat{\mathcal{L}}_{\mathrm{ii}}^{\mathcal{Q}}$ does not play a dynamic role, in which case \textbf{B} may be replaced by the background \textbf{R} or combinations of it, for which $\hat{\mathcal{L}}_{\mathrm{ii}}^{\mathcal{Q}}$ does become volume-invariant in the sense of eq.~(\ref{109}). Note that this type of substitution will not work with Lovelock Lagrangians. To wit, if one of the \textbf{B}'s in $\hat{\mathcal{L}}_{2}$ in eq.~(\ref{4}) were replaced by the corresponding background tensor \textbf{R}, the resulting superpotential would equal only half of $\hat{U}_{2}^{\rho\sigma}$!

One is then tempted to extend the family of Lagrangians to the generalized Palatini formulation, and explore other Lagrangian densities, which couple the Riemann tensor and the non-metricity, and which would not ordinarily produce first-order variational derivatives if the connection and metric were the only independent fields. In each case the question arises: did we make the right choice of variables? All aspects being equal, this matter must be determined by more involved equations that determine the superpotential in the event of higher-order derivatives in the variational derivatives. At present, we may only speculate. One could even envision a geometry in which all possible combinations of non-metric terms are averaged in some fashion (\textquotedblleft path-integral\textquotedblright ) to produce physical observables. Here we limit ourselves to three principal examples:
\begin{eqnarray}
\hat{{\mathcal{L}}}_{\mathrm{I}}^{\mathcal{Q}} & \equiv & \hat{g}_{\rho\sigma}Q_{\kappa }^{~\mu \rho }Q_{\lambda }^{~\nu \sigma }B_{~~~~\mu \nu }^{%
\left[ \kappa \lambda \right] };  \label{138} \\[1ex]
\hat{{\mathcal{L}}}_{\mathrm{II}}^{\mathcal{Q}} & \equiv & \mathfrak{g}^{\rho\sigma }\hat{B}_{\left( \rho \sigma \right) };  \label{139} \\[1ex]
\hat{{\mathcal{L}}}_{\mathrm{III}}^{\mathcal{Q}} & \equiv & \hat{g}_{\rho\sigma}Q_{\kappa }^{~\mu \rho }Q_{\lambda }^{~\nu \sigma }\ast {B}_{~~~\left[ \mu
\nu \right] }^{\kappa \lambda },  \label{140}
\end{eqnarray}
where
\begin{equation}
\mathfrak{g}^{\rho \sigma }\equiv \delta _{\mu \nu }^{\kappa\lambda}Q_{\kappa }^{~\mu \rho }Q_{\lambda }^{~\nu \sigma },~B_{\rho \sigma}\equiv B^{\lambda}_{~\rho \lambda \sigma},~\ast{B}_{~~\mu \nu }^{\kappa \lambda }\equiv \frac{1}{2}B_{\mu \nu \rho\sigma}\eta ^{\rho \sigma \kappa \lambda }.
\label{141}
\end{equation}
$\ast {B}_{~~\mu \nu }^{\kappa \lambda }$ is the right dual of the curvature tensor, and $\eta ^{\rho \sigma \kappa \lambda }$ is the Levi-Civita tensor,
obeying $\varepsilon ^{\rho \sigma \kappa \lambda }=\sqrt{|g|}\eta ^{\rho \sigma \kappa \lambda}$.

A few observations are in order: firstly, note that $\mathfrak{g}^{\left[\rho \sigma \right]}=0$. Secondly, $\hat{{\mathcal{L}}}_{\mathrm{I}}^{\mathcal{Q}}$ and $\hat{{\mathcal{L}}}_{\mathrm{III}}^{\mathcal{Q}}$ are volume-invariant in the sense of eq.~(\ref{109}), and $\hat{{\mathcal{L}}}_{\mathrm{III}}^{\mathcal{Q}}$ reduces to $\hat{{\mathcal{L}}}_{\mathrm{I}}^{\mathcal{Q}}$ for self-dual metrics. Thirdly, $\hat{{\mathcal{L}}}_{\mathrm{III}}^{\mathcal{Q}}$ is entirely first-order, with the metric and affine connection as its dynamic variables, and manifestly violates parity conservation,with explicit dependence on the Levi-Civita tensor density, $\varepsilon^{\mu\nu\rho\sigma}$.

$\hat{{\mathcal{L}}}_{\mathrm{I}}^{\mathcal{Q}}$ and $\hat{{\mathcal{L}}}_{\mathrm{II}}^{\mathcal{Q}}$ represent two possible modifications of $\hat{{\mathcal{L}}}_{1}^{\mathcal{Q}}$, and similarly to $\hat{U}_{\mathrm{ii}}^{\rho\sigma}$, their corresponding superpotentials, $\hat{U}_{\mathrm{I}}^{\rho \sigma}$ and $\hat{U}_{\mathrm{II}}^{\rho \sigma}$ respectively, contribute nothing to the charge at the limit of asymptotic maximal symmetry. To see this, we note that in all three equations (\ref{138})-(\ref{140}) the curvature tensor does not play a dynamic role. It may be replaced by its background analogue without affecting the superpotential on the boundary, at least when \textbf{B} is independent of $\Gamma^{\lambda}_{~\mu\nu}$.

In the case of eqs.~(\ref{138}) and (\ref{139}), the background Riemann and Ricci tensors, respectively, may be decomposed into their irreducible representation, and the scalar curvature terms become proportional to $\hat{{\mathcal{L}}}_{\mathrm{i}}^{\mathcal{Q}}$, for which $\hat{U}_{\mathrm{i}}^{\rho\sigma}=0$ identically (see eq.~(\ref{119})).

In the case of $\hat{{\mathcal{L}}}_{\mathrm{II}}^{\mathcal{Q}}$, this result can be generalized to include a whole class of Lagrangian densities of the form:
\begin{equation}
\hat{{\mathcal{L}}}_{\mathcal{Q}}\left[ g^{\mu \nu },\Gamma_{~\mu\nu}^{\lambda },B_{~\nu \rho \sigma }^{\mu }\right] \equiv \sqrt{\left\vert g\right\vert }\mathfrak{g}^{\rho \sigma }f_{\rho \sigma }\left[ \mathbf{B}\right],
\label{142}
\end{equation}%
where $f_{\rho \sigma }$ is a polynomial function of the curvature tensor. Explicit expressions for $\hat{U}_{\mathrm{I}}^{\rho \sigma }$ and $\hat{U}_{\mathrm{II}}^{\rho \sigma}$ are given in the Appendix. Below we consider $\hat{{\mathcal{L}}}_{\mathrm{III}}^{\mathcal{Q}}$. The variational derivative with respect to the curvature tensor, $\delta _{\mathbf{B}}\hat{{\mathcal{L}}}_{\mathrm{III}}^{\mathcal{Q}}\sim \mathbf{g}\cdot \left( \mathbf{D}\mathbf{g}\right)\cdot\left( \mathbf{D}\mathbf{g}\right) $ does not contribute to the superpotential on the boundary.

The remaining variational derivatives are given by%
\begin{equation}
\frac{\delta \hat{{\mathcal{L}}}_{\mathrm{III}}^{\mathcal{Q}}}{\delta \Gamma
_{~\mu \nu }^{\lambda }}\overset{\bullet }{=}4\hat{g}_{\eta \beta }\delta
_{\lambda }^{\left( \eta \right. }g^{\left. \kappa \right) \left( \mu
\right. }\ast B_{~~~\left[ \kappa \alpha \right] }^{\left. \nu \right)
\sigma }Q_{\sigma }^{~\alpha \beta },
\label{143}
\end{equation}%
and%
\begin{equation}
\frac{\delta \hat{{\mathcal{L}}}_{\mathrm{III}}^{\mathcal{Q}}}{\delta g^{\mu
\nu }}\overset{\bullet }{=}2\hat{g}_{\alpha (\mu }\delta _{\nu )}^{\eta
}\ast B_{~~\left[ \beta \eta \right] }^{\gamma \delta }F_{~~~~\gamma \delta
}^{\left( \alpha \beta \right) }.
\label{144}
\end{equation}%
The superpotential is given by%
\begin{equation}
\begin{split}
\hat{U}_{\mathrm{III}}^{\rho \sigma } = & -\frac{2}{3}\ast {R}_{~~\kappa
\lambda }^{\rho \sigma }\left( \xi ^{\mu }\hat{g}^{\kappa \nu }\Delta _{~\mu
\nu }^{\lambda }+\xi ^{\kappa }\hat{g}^{\mu \nu }\Delta _{~\mu \nu
}^{\lambda }+\xi ^{\kappa }\hat{g}^{\lambda \mu }\Delta _{~\mu \nu }^{\nu
}\right) \\[1ex]
& -\frac{8}{3}\left( \xi ^{\left( \kappa \right. }\hat{g}^{\left. \mu
\right) \left[ \rho \right. }\ast {R}_{~~~\kappa \lambda }^{\left. \sigma %
\right] \nu }-\hat{g}_{\kappa \lambda }\xi ^{\left( \kappa \right.
}g^{\left. \eta \right) \left[ \rho \right. }\ast {R}_{~~~~~\eta }^{\left.
\sigma \right] \mu \nu }\right) \Delta _{~\mu \nu }^{\lambda }.
\end{split}
\label{145}
\end{equation}%
In four dimensions this may be simplified further using the identity%
\begin{equation}
2\delta _{\lambda }^{\left[ \rho \right. }\varepsilon ^{\left. \sigma \right]
\kappa \mu \nu }=-\delta _{\lambda }^{\kappa }\varepsilon ^{\rho \sigma \mu
\nu }-2\varepsilon ^{\rho \sigma \kappa \left[ \mu \right. }\delta _{\lambda
}^{\left. \nu \right] }~\text{for}~N=4.  \label{146}
\end{equation}%
We have%
\begin{equation}
\hat{U}_{\mathrm{III}}^{\rho \sigma }\left( N=4\right) = -\frac{4}{3}\varepsilon ^{\rho \sigma \mu \kappa }\left( \xi ^{\eta
}R_{~\left( \eta \kappa \right) \lambda }^{\nu }\Delta _{~\mu \nu }^{\lambda}+\xi _{\eta }R_{\kappa \lambda }g^{\nu (\eta }\Delta _{~~\mu \nu }^{\lambda
)}\right) .
\label{147}
\end{equation}%
Moreover, in the case ${\bar{C}}_{~\nu \rho \sigma }^{\mu }=0$ and $r_{\mu \nu }=0$, the right-hand side of eq.~(\ref{147}) takes the form
\begin{equation}
\hat{U}_{\mathrm{III}}^{\rho \sigma }\left( N=4\right) =-\frac{1}{9}\varepsilon ^{\rho \sigma \mu \kappa }\bar{g}_{\kappa \lambda }\bar{R}\left(3\bar{\xi}^{[\nu }\Delta_{~\mu \nu}^{\lambda ]}+\bar{\xi}^{(\nu}\Delta_{~\mu \nu}^{\lambda )}\right) .  \label{148}
\end{equation}
Thus $\hat{U}_{\mathrm{III}}^{\rho \sigma}$ possesses a non-zero contribution proportional to the background scalar curvature, from which we learn that non-metricity can have a measurable effect even at the region of asymptotic maximal symmetry.
It is easy to show that the functional terms multiplied by $\bar{R}$ on the r.h.s of eq.~(\ref{148}) originate from
\begin{equation}
{\hat{\mathcal{L}}}^{\mathcal{Q}}_{a}\equiv\varepsilon^{\kappa \lambda \mu \nu} g^{\rho \sigma} Q_{\kappa \mu \rho} Q_{\lambda \nu \sigma}.
\label{148a}
\end{equation}
Specifically, the corresponding superpotential is given by
\begin{equation}
\hat{U}^{\rho \sigma}_{a}\left(N=4\right)=-\frac{4}{3}\varepsilon^{\rho \sigma \mu \kappa}\bar{g}_{\kappa \lambda}\left(3\bar{\xi}^{[\nu }\Delta_{~\mu \nu}^{\lambda ]}+\bar{\xi}^{(\nu}\Delta_{~\mu \nu}^{\lambda )}\right).
\label{148b}
\end{equation}

\subsection{Example: Asymptotic AdS Solutions}

All the parity-preserving Lagrangian densities with explicit dependence on the non-metricity we have studied in this section produced superpotentials that depended solely on the background Weyl and trace-free Ricci tensors. If these superpotentials can contribute anything on the boundary, it would be in a space-time that is either not asymptotically conformally flat or with sources at the background. The latter violates the very definition of the concept of isolated and localized sources, while the former is linked to anisotropy of the background. Space-times with local anisotropy exhibit a certain scale dependence at every point of space-time, and as far as we know it is in itself a consequence of matter. This means that such geometries do not describe strictly isolated sources of gravity.

It has already been argued that non-metricity effects can be induced by matter \cite{SoL}. $\hat{{\mathcal{L}}}_{\mathrm{III}}^{\mathcal{Q}}$ is parity-violating, coupling non-metricity with the curvature tensor, and as such is an example of a Chern-Simons modified gravity. The field equations derived from this density are trivially satisfied by any metric and its associated metric-compatible connection. The resulting superpotential, $\hat{U}_{\mathrm{III}}^{\rho \sigma}$, affords us an opportunity to examine whether non-metricity emanating from a source of gravity could have a measurable effect at the boundary, which is torsion-less, metric-compatible, source-free and maximally-symmetric.

To this end, we examine two metrics, which, at the absence of sources are asymptotically (locally) AdS in four dimensions. The first one is due to Henneaux and Teitelboim \cite{HT}. It describes the asymptotic behaviour of anti-de Sitter space-time with a cosmological constant, mass and angular momentum, and is given by the non-zero components of the metric in spherical coordinates:
\begin{subequations}
\begin{eqnarray}
g_{tt}& = &-\frac{r^{2}}{l^{2}}+\frac{2M\lambda ^{5}}{r}+O\left( \frac{1}{r^{3}%
}\right) ;~g_{t\phi } = -\frac{2J\lambda ^{5}\sin ^{2}{\theta }}{r}+O\left( \frac{1}{%
r^{3}}\right) ;\\[1ex]
g_{\phi \phi }& = & r^{2}\sin ^{2}\theta \left( 1+\frac{2Ma^{2}\lambda
^{5}\sin^{2}\theta }{r^{3}}\right) +O\left( \frac{1}{r^{3}}\right) ; \\[1ex]
g_{rr}& = &\frac{l^{2}}{r^{2}}-\frac{l^{4}}{r^{4}}+\frac{2Ml^{4}\lambda ^{3}}{%
r^{5}}+O\left( \frac{1}{r^{7}}\right) ; \\[1ex]
g_{\theta r}& = & -\frac{Ml^{2}a^{2}\lambda ^{5}\sin 2\theta }{r^{4}}+O\left(%
\frac{1}{r^{6}}\right) ;~g_{\theta \theta} = r^{2}+\frac{Ma^{4}\lambda ^{7}\sin ^{2}\left(2\theta\right) }{2r^{3}}+O\left( \frac{1}{r^{5}}\right),
\end{eqnarray}
\label{149}
\end{subequations}
where
\begin{equation}
\alpha \equiv \frac{a}{l};~\lambda \equiv \frac{1}{\sqrt{1-\alpha^{2}\sin^{2}\theta }};~J\equiv aM.
\label{150}
\end{equation}
The background metric is obtained by taking the sources to vanish, given by the line element in the regime $r\gg l$:%
\begin{equation}
d\bar{s}^{2}\approx -\frac{r^{2}}{l^{2}}dt^{2}+\left( \frac{l^{2}}{r^{2}}-\frac{l^{4}}{r^{4}}\right) dr^{2}+r^{2}\left( d\theta ^{2}+\sin ^{2}\theta d\phi ^{2}\right).
\label{151}
\end{equation}
The question then becomes, can $\hat{U}_{\mathrm{III}}^{\rho \sigma }$ in eq.~(\ref{148}) contribute to the mass or angular momentum. In four dimensions $dS_{\rho \sigma }=\frac{1}{2}\epsilon _{\rho \sigma \mu \nu }dx^{\mu }\wedge dx^{\nu }$, so the integrand in eq.~(\ref{66}) is given by
\begin{equation}
\hat{U}_{\mathrm{III}}^{\rho \sigma}dS_{\rho \sigma} = \frac{2}{9}dx^{\mu}\wedge dx^{\kappa }\bar{g}_{\kappa \lambda }\bar{R}\left( 3\bar{\xi}^{[\nu }\Delta _{~\mu \nu }^{\lambda]}+\bar{\xi}^{(\nu }\Delta _{~\mu\nu }^{\lambda )}\right) ,
\label{152}
\end{equation}
with
\begin{equation}
\bar{R}=4\Lambda~\text{and}~\Lambda =-\frac{3}{l^{2}}.
\label{153}
\end{equation}
The closed two-surface at spatial infinity is a $2$-sphere, $S_{2}$, defined by constant $r$ and $t$, \emph{i.e.} $dr=dt=0$ at the boundary. For the
asymptotically time-like Killing vector field $\bar{\xi}^{\mu}=\left(1,0,0,0 \right)$, we have
\begin{equation}
\mathcal{E}\equiv \frac{4}{9}\bar{R}\bar{\xi^{t}}\lim_{r\rightarrow
\infty}\int {d\theta d\phi \left( \bar{g}_{\phi \phi }\Delta _{~t\theta
}^{\phi }-\bar{g}_{\theta \theta }\Delta _{~t\phi }^{\theta }\right) }.
\label{154}
\end{equation}%
A direct calculation using the metric in eq.~(\ref{149}) reveals $\mathcal{E}=0$, since $\Delta _{~t\theta }^{\phi },\Delta _{~t\phi }^{\theta }\propto {r^{-3}}$, so that $\hat{U}_{\mathrm{III}}^{\rho \sigma }dS_{\rho\sigma}=O\left( \frac{1}{r}\right) $, which vanishes in the limit $%
r\rightarrow \infty $. This means that $\hat{U}_{\mathrm{III}}^{\rho \sigma }$ does not contribute to the energy in that geometry. For the asymptotically axi-symmetric Killing vector field $\bar{\xi}^{\mu}=\left(0,0,0,1 \right)$ we have
\begin{equation}
\mathcal{J}\equiv -\frac{2}{9}\bar{R}\bar{\xi}^{\phi}\lim_{r\rightarrow\infty }\int {d\theta d\phi \left( \bar{g}_{\phi \phi }\Delta _{~\theta \nu}^{\nu }+2\bar{g}_{\theta \theta }\Delta _{~\phi \phi }^{\theta }\right)}.
\label{155}
\end{equation}%
Again, from eq.~(\ref{149}) it follows that $\Delta_{~\theta \nu}^{\nu},\Delta_{~\phi \phi }^{\theta }\propto r^{-3}$, so $\mathcal{J}=0$ as well.

It would appear that the typical asymptotically AdS structure does not yield contributions from such terms. An example that does produce a non-vanishing contribution to the angular momentum is inspired by the Taub-NUT metric, and its derivation \cite{LBNZ} (see also \cite{B}). We start with the well-known split of the stationary metric as described by Landau and Lifshitz \cite{LL}%
:
\begin{equation}
ds^{2}=-e^{-2\nu }\left( dt-A_{i}dx^{i}\right) ^{2}+g_{ij}^{\left(3\right)}dx^{i}dx^{j},
\label{156}
\end{equation}%
where $\nu,A_{i},g_{ij}^{\left( 3\right) }$ are independent of $t$, and here the lower case Latin indices enumerate the spatial part of the metric, \emph{i.e.} $i,j=\{1,2,3\}=\{r,\theta ,\phi \}$. Owing to the gauge freedom in $t$, in analogy with classical electromagnetism, we define the \emph{gravo-electric} and \emph{gravo-magnetic} fields by
\begin{equation}
\mathbf{E}_{g}\equiv \overrightarrow{\nabla }\nu ,~\mathbf{B}_{g}\equiv
\overrightarrow{\nabla }\times A,  \label{157}
\end{equation}%
respectively, where the differential operator, $\overrightarrow{\nabla}$, is with respect to $g_{ij}^{\left( 3\right) }$. It was shown \cite{LBNZ} that the choice
\begin{equation}
\nu \left( r\right) =-\frac{1}{2}\ln \left[ 1-\frac{2}{r^{2}}\left( q^{2}+m%
\sqrt{r^{2}-q^{2}}\right) \right] ;~ A =\left( 0,0,A_{\phi }\right) ~\text{with}~A_{\phi }\left( \theta \right)=2q\left( 1+\cos \theta \right) ,
\label{158}
\end{equation}
together with the metric components%
\begin{equation}
g_{rr}^{\left( 3\right) }=e^{2\nu }\left( 1-\frac{q^{2}}{r^{2}}%
\right)^{-1},~g_{\theta \theta }^{\left( 3\right) }=r^{2},~g_{\phi \phi
}^{\left( 3\right) }=r^{2}\sin ^{2}\theta ,  \label{160}
\end{equation}%
solves Einstein's field equations in vacuo, and reproduces the Taub-NUT metric. This beautiful analogy with classical electromagnetism lends itself naturally to other extensions. In particular, we can define:
\begin{equation}
\nu \left( r\right) = -\frac{1}{2}\ln \left( 1+\frac{r^{2}}{l^{2}}\right);
~A = \left( 0,0,A_{\phi }\right) ~\text{with}~A_{\phi } \left(\theta ,r  \right)= \frac{2q^{2}\lambda\left(\theta\right)}{r},
\label{161}
\end{equation}
where we have assumed for simplicity the massless source approximation \cite{BONNOR},\emph{i.e.} $m=0$, at the outset. The line element is given by:
\begin{equation}
ds^{2}= -\left(1+\frac{r^{2}}{l^{2}}\right)\left(dt-\frac{2q^2\lambda(\theta)}{r}d\phi\right)^{2}
+\left(1+\frac{r^{2}}{l^{2}}\right)^{-1}dr^{2} +r^{2}d\theta ^{2}+ r^{2}\sin ^{2}\theta d\phi ^{2}.
\label{163}
\end{equation}
Eq.~(\ref{163}) reduces to eq.~(\ref{151}) when $q=0$ for large $r$. The latter is taken to be the background in this calculation.

A simple calculation shows that $\Delta _{~\theta \nu }^{\nu }=0$, while $\Delta_{~\phi \phi }^{\theta }=2 n^{4}\frac{l^2}{r^2}\frac{d}{d\theta}\lambda^2\left(\theta\right)+O\left(\frac{1}{r^{4}}\right) $. It follows that only the second term in the integrand in eq.~(\ref{155}) can contribute.  In particular, if $\lambda\left( \theta\right)=\frac{1}{2}\sin\left(\theta/2\right)$, we find
\begin{equation}
\mathcal{J}=-\frac{4}{9}\bar{R}\lim_{r\rightarrow \infty}\int_{0}^{\pi }{d\theta \int_{0}^{2\pi }d\phi \bar{g}_{\theta \theta}\Delta _{~\phi \phi }^{\theta }}=\frac{16\pi }{3}n^{4},
\label{164}
\end{equation}
where $n\equiv q/l$. Hence the gravo-magnetic charge in this model couples to the scale $l$, with a contribution of the order of $l^{-4}$.

It is interesting to note that the calculation of the energy from this superpotential results in a divergent quantity. Indeed, since $\Delta _{~t\phi}^{\theta}=-\frac{n^{2}\lambda^{\prime}(\theta)}{r}+O\left( \frac{1}{r^3}\right)$ and $\Delta_{~t\theta}^{\phi}=\frac{n^{2}\lambda^{\prime}(\theta)}{r\sin^{2}\theta}+O\left( \frac{1}{r^3}\right)$, we find for the aforementioned $\lambda$ (see eq.~(\ref{154})):
\begin{equation}
\mathcal{E}= \frac{4}{9}n^{2}\bar{R}\lim_{r\rightarrow \infty} \left( 2\pi r \right).
\label{165}
\end{equation}%
Not only is $\mathcal{E}$ infinite, but is negative to boot! This is akin to the singularity along the half-axis $\theta =\pi$ in the Taub-NUT metric in four dimensions. Unlike Misner \cite{Mis}, who removed this singularity by re-identifying $t$ as a periodic coordinate, Bonnor \cite{BONNOR} has considered the alternative of retaining the singularity and a causal structure for $t$, and has suggested the singularity be interpreted as a \textquotedblleft massless source of angular momentum.\textquotedblright\ Israel \cite{We} pointed out that the Bonnor singularity was not a line source in the typical sense, since its circumference is different from zero. Presumably it is this which makes a non-vanishing angular momentum possible for this source. Indeed, $\mathcal{E}$ is proportional to the circumference, and the limit $r\rightarrow\infty$ implies it is expanded beyond the z-axis, suggesting that the source is not strictly isolated.

A few remarks are in order: (a) the metric in eq.~(\ref{163}) does not solve Einstein's vacuum field equations. In fact, the scalar curvature and the Kretschmann scalar are given by
\begin{eqnarray}
R&=&\bar{R}+\frac{2n^4\lambda^2(\theta)}{r^2\sin^{2}\theta}+O\left(\frac{1}{r^4}\right);\label{166}\\[1ex]
K&=&R^{\mu \nu\rho \sigma}R_{\mu \nu\rho \sigma}=\frac{24}{l^4}-\frac{24n^4\lambda^2\left(\theta\right)}{l^2\sin^2{\theta}}\frac{1}{r^2}+O\left(\frac{1}{r^4}\right), \label{167}
\end{eqnarray}
respectively, and it can be shown that $G_{\mu \nu}+\Lambda g_{\mu \nu}=qT_{\mu \nu}$, so that with $q=0$ we revert back to the AdS background. (b) Since this metric is only asymptotically locally AdS, a more detailed analysis of the asymptotic structure of the metric and its singularities is necessary in order to determine its background behaviour. In particular, $\mathcal{E}$ and $\mathcal{J}$ in eqs.~(\ref{164})-(\ref{165}) originate from the dominant contributions, $g_{t\phi}\approx2{n^2}\lambda\left(\theta\right)r$ and $g_{\phi \phi}\approx-4q^{4}l^{-2}\lambda^{2}\left(\theta\right)$ respectively. In particular, if $\lambda\left( \theta\right)=\frac{1}{2}\sin\left(\theta/2\right)$, then from eq.~(\ref{166}) it follows that there is a singularity at $\theta=\pi$, and $g_{\phi \phi}$ cannot be neglected.

\section{\label{sec:level5}Summary and Discussion}
In this paper we set out to explore the inverse problem of superpotentials, \emph{i.e.}, the derivation of the flux from the current. We formulated the problem using two distinct formulations of metric-affine gravity, distinguished by their treatment of the non-coordinate base (soldering form), $\theta _{~\mu }^{a}$. One is the general linear formulation, in which $\theta _{~\mu }^{a}=\theta _{~\mu }^{a}\left( x^{\lambda }\right)$ is a dynamic variable to be determined by the field equations on the boundary. It was not constrained beyond the requirement of its invertibility. At the boundary it was fixed by the canonical (metric) choice: $\left. \theta_{~\mu}^{a}\right\vert_{\partial \mathcal{M}}=\delta_{~\mu }^{a}$. The other is the Palatini formulation, in which the non-coordinate base had been fixed already at the very start. This subtle difference had dictated everything that followed.

In the general linear formulation, this meant that two additional independent dynamic variables could be constructed, the fibre metric and the spin connection, both of which behave as perfect tensors under diffeomorphism. This guaranteed that variational symmetries of the fields would contain no higher than first-order derivatives of the fields and their symmetry parameters. Together with the first-order variational derivatives from Lovelock Lagrangians, this formulation produced a simple and elegant equation for the current, expressed as a linear combination of variational derivatives, from which the superpotential was uniquely determined.

However, the same cannot be said for the Palatini formulation. Forcing $\theta _{~\mu }^{a}\left( x^{\lambda }\right) =\delta _{~\mu }^{a}$ everywhere in $\mathcal{M}$ meant that it was no longer a viable independent dynamic variable. Now only two variables remained, the space-time metric $g^{\mu \nu}$, and the torsion-less affine connection, $\Gamma_{~\mu\nu}^{\lambda}$; and while the variational derivatives still contained no higher than first-order derivatives, the second-order derivatives associated with $\pounds _{\xi }\Gamma_{~\mu \nu }^{\lambda}$ meant it was not possible to use the simpler general linear prescription. It was, however, still possible to solve the series of differential identities, and to obtain the current, and subsequently, the superpotential. A major difference between the formulations was the presence of second-order derivatives in the conserved current. This suggested that more boundary conditions were needed to obtain a unique superpotential. Consequently, the current was no longer given by a simple linear combination of variational derivatives, but was more complicated, with derivatives of the fields and their variational derivatives.

One major finding in this paper is the fact that the current and superpotential obtained from the Palatini formulation are more general, encompassing the results of the general linear formulation as particular cases. Moreover, we have found that the same current can induce two distinct superpotentials within the Palatini formulation, in contrast to the unique superpotential obtained in the general linear formulation.

This led us to a paradox, stemming from the fact that the dynamics on the boundary are identical. Indeed, we have shown explicitly that the dynamics imposed by the fixation of $\theta_{~\mu }^{a}\left( x^{\lambda }\right)$ at the boundary, and the metricity condition, $\left. D_{\lambda }g^{\mu \nu}\right\vert _{\partial \mathcal{M}}=0$, are equivalent for both formulations. How is it, then, possible that their respective currents and superpotentials differ? Surely the mass of an isolated star far far away is not privy to our choice of fields or dynamic variables!

We have presented two solutions to this paradox. One solution required that space-time be asymptotically maximally symmetric. At first glance this condition may appear too restrictive. Indeed, why must space-time possess such a global topology? Nonetheless, we have given several arguments in its favour, centered on the idea, that the Weyl and trace-free Ricci tensors do not vanish for space-times with non-localized sources, that are present at the boundary as well. Hence, the requirement of asymptotic maximal symmetry may turn out to be equivalent to the requirement for the localization of isolated charges. Therefore, it is not surprising that we have not been able to find a solution to the second-order Lovelock Lagrangian in the literature,
describing isolated sources of gravity, which is not asymptotically maximally symmetric. That, however, does not mean that such solutions do not exist.

Our second solution does not impose additional constraints on the global topology of space-time. Instead, we have shown that in order to resolve this paradox in a manner consistent with our integrability conditions, the Lagrangian must be renormalized. In this case the renormalized second-order Lovelock Lagrangian density is given by
\begin{equation}
\hat{\mathcal{L}}_{2}^{\prime \mathcal{P}}=\frac{1}{4}\delta _{\alpha \beta \gamma \delta }^{\mu \nu \rho \sigma }\left( \hat{B}_{~~\mu \nu }^{\alpha \beta }-\hat{g}_{\kappa \lambda }Q_{\mu }^{~\alpha \kappa }Q_{\nu }^{~\beta \lambda}\right) B_{~~\rho \sigma }^{\gamma \delta }.  \nonumber
\end{equation}
Its extension to higher orders of curvature is straightforward.

We resolved this paradox, but it only served to highlight the new possibilities of non-metricity. Working within the Palatini formulation, we have shown that non-metricity in the bulk of space-time can be used to induce new terms in the superpotential. In particular, we have shown how a parity-violating action with explicit coupling between the curvature and quadratic non-metricity can induce conserved charges on the boundary, and consequently that it can influence measurements at the asymptotically maximally symmetric regime. This effect, however, is not limited to maximal symmetry. In fact, we have given an example of a Lagrangian density, ${\hat{\mathcal{L}}}^{\mathcal{Q}}_{a}=\varepsilon^{\kappa \lambda \mu \nu} g^{\rho \sigma} Q_{\kappa \mu \rho} Q_{\lambda \nu \sigma}$, with no dependence at all on the curvature, which produces a viable superpotential.

This seems to fit with Percacci \cite{P}, who has already noted that quadratic terms in non-metricity appear as part of the gravitational Higgs phenomena. While the original Higgs mechanism \cite{EB,Higgs,GHK} in QFT was used exclusively to generate mass or energy, the present realization allows for a broader class of asymptotic symmetries for a given geometry. In our toy model, we have calculated a finite contribution to the angular momentum arising from the coupling of the gravo-magnetic charge to the cosmological scale.

Moreover, we have shown that quadratic non-metricity can be used to complement the surface term proposed by Obukhov \cite{Obukhov}. In fact, the renormalized Lagrangian density $\hat{\mathcal{L}}_{1}^{\prime \mathcal{P}}$ fixed the variation of this term on the boundary, and since $\hat{U}_\mathrm{i}^{\rho \sigma}=0$, the introduction of this surface term to the action allowed us to determine the proper coupling of the non-metricity terms to $\hat{\mathcal{L}}_{1}^{\mathcal{P}}$.

Exirifard and Sheikh-Jabbari take the standpoint\cite{ESJ} that: \textquotedblleft the physically allowed theories of the modified gravity are the ones for which the Palatini and the metric formulations are (classically) equivalent\textquotedblright. Furthermore, they claim that Lovelock Lagrangians satisfy this requirement, especially in asymptotically flat space-times, and should thus be considered as those plausible modifications.

Our results show that even within the first-order formulation, Lovelock Lagrangians must be treated with considerable care. Our results seem to support their claim in as much as asymptotic flatness is such that there are no sources at infinity and the background Weyl tensor vanishes identically. However, our result appears to indicate that even the extended Lovelock family, with quadratic curvature terms, can lead us away from purely metric results.

\section{\label{sec:level6}Appendix: Detailed Calculations}

In this short Appendix, we provide several extensions and elucidations to the text.

In section \textrm{III}.B, we proclaimed three boundary conditions in eq.~(\ref{86}). In addition to the vanishing of non-metricity, we demanded the vanishing of its variation as well. This condition is not, strictly speaking, a new condition. It is consistent with the requirement that the connection be reduced to the Christoffel symbol on the boundary. Let us show this explicitly below. On the one hand, we have:
\begin{equation}
\delta Q_{\lambda}^{~\mu \nu} = D_{\lambda} \delta g^{\mu \nu} + 2 g^{\kappa \left( \mu \right.} \delta \Gamma^{\left. \nu \right)}_{~\kappa \lambda}.
\label{A.1}
\end{equation}
On the other hand, we require that on the boundary the connection be metric-compatible, which in turn implies that
\begin{equation}
\left. \delta \Gamma^{\rho}_{~\kappa \lambda} \right\vert_{\partial \mathcal{M}} = \frac{1}{2} g^{\rho \sigma} \left(  D_{\kappa} \delta g_{\lambda \sigma} + D_{\lambda} \delta g_{\kappa \sigma} - D_{\sigma} \delta g_{\kappa \lambda} \right) .
\label{A.2}
\end{equation}
Substituting eq.~(\ref{A.2}) into the second term in the r.h.s. of eq.~(\ref{A.1}), we find on the boundary
\begin{equation}
\left. \delta Q_{\lambda}^{~\mu \nu} \right\vert_{\partial \mathcal{M}} = -2 g^{\rho \left(\mu \right.}Q_{\lambda}^{~\left.\nu\right) \sigma} \delta g_{\rho \sigma}.
\label{A.3}
\end{equation}
The l.h.s. of eq.~(\ref{A.3}) vanishes on the boundary, since either $\left. Q_{\lambda}^{~\mu \nu} \right\vert_{\partial \mathcal{M}}=0$ or $\left. \delta g_{\rho \sigma}\right\vert_{\partial \mathcal{M}} = 0$.

$\mathbf{Z}$-calculation details: In section \textrm{III}.C we introduced a re-shuffling of the Noether density. We claimed that the anti-symmetric density $\hat{Z}^{\rho \sigma }$ defined in eq.~(\ref{100}), and which appears in eq.~(\ref{102}), did not contribute to the superpotential in eq.~(\ref{107}). Here is the proof for $n=2$ discussed in the main text. On the one hand we find
\begin{equation}
2g^{\kappa \left( \mu \right. }\delta _{\eta }^{\left. \nu \right) }\frac{\partial \hat{Z}^{\rho \sigma }}{\partial \partial _{\lambda }g^{\mu\nu }} \delta \Gamma_{~\kappa \lambda}^{\eta } = 2\xi ^{\alpha }\delta _{\alpha \beta \gamma \delta }^{\lambda \left[ \rho\right\vert \eta \tau }B_{~\epsilon \eta \tau }^{\gamma }g^{\kappa \mu }\hat{h}_{\mu \nu }^{\beta \left\vert \sigma \right] \delta \epsilon}\delta \Gamma _{~\kappa \lambda }^{\nu }.
\label{A.4}
\end{equation}
While on the other hand:
\begin{equation}
-\frac{\partial \hat{Z}^{\rho \sigma }}{\partial \Gamma_{~\mu \nu}^{\lambda }}\delta \Gamma _{~\mu \nu }^{\lambda }= -2\xi ^{\alpha }\delta _{\alpha \beta \gamma \delta }^{\lambda \left[ \rho\right\vert \eta \tau }B_{~\epsilon \eta \tau }^{\gamma }g^{\kappa \mu }\hat{h}_{\mu \nu }^{\beta \left\vert \sigma \right] \delta \epsilon}\delta \Gamma _{~\kappa \lambda }^{\nu }.
\label{A.5}
\end{equation}
So they cancel each other out. It is easy to show that it is valid for all the Lovelock Lagrangians in general.

In section IV.A, eq.~(\ref{110}) we defined $\hat{\mathcal{L}}_{\mathrm{ii}}^{\mathcal{Q}}$. Its extension to higher orders is straightforward. For example, below is the third order, $\hat{\mathcal{L}}_{\mathrm{iii}}^{\mathcal{Q}}$:
\begin{equation}
\hat{\mathcal{L}}_{\mathrm{iii}}^{\mathcal{Q}}=\hat{g}_{\rho \sigma }\delta_{\kappa \lambda \alpha \beta \theta \phi }^{\mu \nu \gamma \delta\varepsilon \eta }B_{~~\varepsilon \eta }^{\theta \phi }B_{~~\gamma \delta}^{\alpha \beta }Q_{\mu }^{~\kappa \rho }Q_{\nu }^{~\lambda \sigma }.
\label{A.8}
\end{equation}

The algorithm is clear: each additional curvature tensor is coupled to the previous order by adding two covariant and two contravariant indices to the generalized Kronecker delta. It should be noted that in $N\geq 6$ dimensions it is possible to form additional parity-preserving Lagrangians from the non-metricity and the curvature tensors which produce first-order field equations. For example, such Lagrangians may have combinations of four non-metricity tensors coupled to a single curvature tensor (in six dimensions), etc.

The symmetry discussed in section \textrm{IV}.B is relevant here. Any additional curvature terms that are coupled  in this manner to $\hat{\mathcal{L}}_{\mathrm{ii}}^{\mathcal{Q}}$ are essentially background quantities, contributing to the superpotential on the boundary as a numerical pre-factor. Therefore, the factor $2$ that appears in eq.~(\ref{122}) - which results from the symmetry of the variation of the non-metricity terms - remains constant. Consequently, for $n>2$ a different normalization constant is required, which depends on the order. For example, for $n=3$ the re-normalized density $\hat{\mathcal{L}}_{3}^{\prime \mathcal{P}}$ reads
\begin{equation}
\hat{\mathcal{L}}_{3}^{\prime \mathcal{P}}=\frac{1}{8}\delta^{\kappa \lambda \mu \nu \rho \sigma}_{\alpha \beta \gamma \delta \epsilon \eta}\left(B^{\alpha \beta}_{~~\kappa \lambda}-\frac{3}{2}g_{\tau \chi}Q_{\kappa}^{~\alpha \tau }Q_{\lambda}^{~\beta\chi} \right)B^{\gamma \delta}_{~~\mu \nu}B^{\epsilon \eta}_{~~\rho \sigma}.
\end{equation}
In general, the pre-factor of the non-metricity terms in parenthesis is $-n/2$. It follows that the difference between the KL and JS superpotentials for $n$th ordered Lovelock Lagrangian, $\Delta \hat{U}_{n}^{\rho \sigma}$, is constructed on top of the $n=2$ structure, which explains the absence of the scalar curvature in higher orders as well.

In section \textrm{IV}.B we discussed the parity-preserving Lagrangian densities, stating that the superpotentials they induced on the boundary did not contain any scalar curvature terms, when expressed in terms of the irreducible background tensors. Below we give the expressions for $\hat{U}_{\mathrm{I}}^{\rho \sigma }$ and $\hat{U}_{\mathrm{II}}^{\rho \sigma }$, which correspond to the Lagrangian densities in eqs.~(\ref{138}) and (\ref{139}) respectively:
\begin{equation}
\hat{U}_{\mathrm{I}}^{\rho \sigma }=-\frac{4}{3}\bar{\hat{\xi}}^{\kappa }\Delta^{\lambda}_{~\mu \nu}\left( \bar{g}^{\mu \left[ \rho \right.}R_{~~~\kappa \lambda }^{\left. \sigma \right] \nu }-\delta _{\lambda }^{\left[ \rho \right. }R_{~~~~~\kappa}^{\left. \sigma \right] \mu \nu } +R^{\rho \sigma}_{~~\kappa \eta} \bar{g}^{\nu(\mu}\delta^{\eta)}_{\lambda} \right),
\label{A.6}
\end{equation}
and
\begin{equation}
\hat{U}_{\mathrm{II}}^{\rho \sigma }= -\frac{8}{3}\bar{\hat{\xi}}^{\left[\rho \right. }\delta _{\kappa }^{\left. \sigma \right] }\left( \bar{g}^{\mu \left[ \kappa \right. }\bar{R}_{~\lambda }^{\left. \nu \right] }+\delta_{\lambda }^{\left[ \kappa \right. }\bar{R}^{\left. \nu \right] \mu }\right)\Delta _{~\mu \nu }^{\lambda } -\frac{8}{3}\bar{\xi}^{\mu }\bar{\hat{g}}^{\nu \left[ \rho \right. }\bar{R}^{\left.\sigma \right] \lambda }\bar{g}_{\kappa \left( \lambda \right. }\Delta_{~\left. \nu \right) \mu }^{\kappa }.
\label{A.7}
\end{equation}
Substituting the background irreducible tensors from eq.~(\ref{73}) into eqs.~(\ref{A.6}) and (\ref{A.7}) one finds the scalar curvature terms vanish, as expected.

\begin{acknowledgments}
The author thanks Joseph Katz for introducing him to the field of asymptotic symmetries and conserved charges, and to some of the open questions in GR. His guidance and support have inspired and encouraged this work. The author acknowledges financial support from the Institute of Chemistry and from Danny Porath during his PhD in experimental molecular Nano-Science. This work is dedicated to Leyb Shteyngarts.
\end{acknowledgments}

\end{document}